\documentclass[twocolumn,superscriptaddress,epsf,psfig]{revtex4-2}
\usepackage{amssymb}
\usepackage{epsfig}
\DeclareMathAlphabet{\pazocal}{OMS}{zplm}{m}{n}            
\usepackage{graphicx}
\usepackage{color}
\usepackage{bm}
\usepackage[normalem]{ulem}     
\usepackage{soul}
\usepackage{hyperref}
\usepackage{amsmath}
\usepackage{braket} 
\usepackage{rotating}
\usepackage{multirow} 
\usepackage[dvipsnames]{xcolor}
\newcommand{\Cro}{Cr$_2$O$_3$}
\newcommand{\Feo}{Fe$_2$O$_3$}
\newcommand{\aFeo}{$\alpha$-Fe$_2$O$_3$}

\begin{document}

\preprint{APS/123-QED}

\title{
Non-relativistic ferromagnetotriakontadipolar order and spin splitting in hematite \\
}

\author{X. H. Verbeek$^\dag$}
 \email{xverbeek@ethz.ch}
 \affiliation{Materials Theory, ETH Zurich, Wolfgang-Pauli-Strasse 27, 8093 Zurich, Switzerland}
\author{D. Voderholzer}
\thanks{These authors contributed equally}
\affiliation{Materials Theory, ETH Zurich, Wolfgang-Pauli-Strasse 27, 8093 Zurich, Switzerland} 
\author{S. Sch\"aren}
\affiliation{Materials Theory, ETH Zurich, Wolfgang-Pauli-Strasse 27, 8093 Zurich, Switzerland} 
\author{Y. Gachnang}
\affiliation{Materials Theory, ETH Zurich, Wolfgang-Pauli-Strasse 27, 8093 Zurich, Switzerland} 
\author{N. A. Spaldin}
\affiliation{Materials Theory, ETH Zurich, Wolfgang-Pauli-Strasse 27, 8093 Zurich, Switzerland} 
\author{S. Bhowal}
\affiliation{Materials Theory, ETH Zurich, Wolfgang-Pauli-Strasse 27, 8093 Zurich, Switzerland} 
\affiliation{Department of Physics, Indian Institute of Technology Bombay, Mumbai 400076, India}

\date{\today}

\begin{abstract}
We show that hematite, $\alpha$-Fe$_2$O$_3$, below its Morin transition, has a ferroic ordering of rank-5 magnetic triakontadipoles on the Fe ions. In the absence of spin-orbit coupling, these are the lowest-order ferroically aligned magnetic multipoles, and they give rise to the $g$-wave non-relativistic spin splitting in hematite. We find that the ferroically ordered magnetic triakontadipoles result from the simultaneous antiferroic ordering of the charge hexadecapoles and the magnetic dipoles, providing a route to manipulating the magnitude and the sign of the magnetic triakontadipoles as well as the spin splitting. Furthermore, we find that both the ferroic ordering of the magnetic triakontadipoles and many of the spin-split features persist in the weak ferromagnetic phase above the Morin transition temperature.  
\end{abstract}
\maketitle
\section{Introduction}

Recently, there has been a surge of interest in an unconventional spin splitting observed in the band structure of collinear compensated antiferromagnetic (AFM) materials. The key feature of this class of AFM materials, often referred to as `altermagnets', is the large spin splitting they exhibit, surpassing typical Rashba splitting magnitudes without requiring spin-orbit coupling (SOC) \cite{PekarRashba1964, Rashba1960, BychkovRashba}.
The interest in this class of antiferromagnets originates from the intriguing symmetries underlying the deviation from typical degenerate spin-polarized bands in conventional antiferromagnets \cite{Libor2020, Naka2019, Kyo-Hoon2019, Hayami2019, Yuan2020, Yuan2021, Smejkal2022PRX, yuan_degeneracy_2023, yuan_uncovering_2023}, and the potential for offering exotic physics that results from it, including AFM spintronics \cite{Smejkal2022, Smejkal2022PRX, Hernandez2021, Shao2021, Bose2022, Bai2022, Karube2021, Libor2020, Helena2020, Feng2020, Betancourt2021, SmejkalAHE2022, Libor2023}, giant magnetoresistance \cite{Libor2022}, chiral magnons \cite{Libor2023}, and superconductivity \cite{Mazin2022, Zhu2023, Zhang2024}.

The non-relativistic spin splitting (NRSS) in collinear compensated antiferromagnets requires broken {\it global} time-reversal symmetry, as well as a specific correlation between the structural symmetry of the magnetic sub-lattices, dictated by the surrounding non-magnetic atomic environment, and the specific arrangement of the anti-parallel magnetic dipolar ordering. The absence of time-reversal symmetry in an antiferromagnet signifies the presence of a ferroic ordering of higher-order magnetic multipoles, which can lead to NRSS within the spin-polarized bands, akin to the trivial NRSS in ferromagnets due to the conventional ferromagnetic (FM) dipolar ordering. Indeed, previous studies have demonstrated that in AFM materials exhibiting a $d$-wave pattern of NRSS, characterized by two nodal planes ($l=2$) of degenerate spin-polarized bands, there exists a ferroic ordering of rank-3 inversion-symmetric and time-reversal-odd magnetic octupoles \cite{BhowalSpaldin2024}. Identifying such higher-order ferroic magnetic multipoles provides insights into the nature of NRSS and its tunability, and contributes to a broader understanding of the physical properties of these unconventional antiferromagnets \cite{BhowalSpaldin2024}. 

Notably, the $d$-wave NRSS pattern is only one among several other patterns predicted and observed so far. For instance, the $g$-wave pattern, characterized by four nodal planes ($l=4$), has recently garnered considerable attention \cite{vanderLaan2023,Lee2024,Krempask2024, reimers_direct_2024}. However, the magnetic octupoles responsible for the $d$-wave spin splitting, can not explain $g$-wave spin splitting due to their quadratic spatial dependence. Identifying the corresponding ferroic magnetic multipole in $g$-wave spin-split antiferromagnets is the topic of this work.

Taking magnetic hematite (\aFeo{}) as an $g$-wave altermagnetic material \cite{Smejkal2022PRX}, we demonstrate that the magnetic ground state is a ferroic ordering of rank-5 magnetic triakontadipoles. These magnetic triakontadipoles exist without spin-orbit interaction, and they form the lowest-order ferroically ordered magnetic multipole in the absence of SOC. Our calculations reveal a correlation between the magnetic triakontadipoles on the Fe ions and their local coordination environment, characterized by charge hexadecapoles, suggesting the manipulation of the magnetic triakontadipole by controlling the positions of the oxygens surrounding the Fe ions. Such a manipulation offers in turn a means to regulate both the magnitude and sign of the spin splitting in hematite. Our study therefore provides both a multipolar description of the $g$-wave spin splitting and a framework for controlling it via the magnetic triakontadipole.

The remainder of this manuscript is structured as follows. In Section \ref{methods}, we review the crystal structure and magnetic ground state of $\alpha$-Fe$_2$O$_3$, and summarize the computational techniques used in this study. We present the results of our electronic structure calculations of the charge and magnetic multipoles and their relationship to the spin-splitting in section \ref{results}. In addition, we explore methods for tuning the magnetic triakontadipole and the spin splitting by modifying the magnetic ordering and structure, as well as changes in properties above the Morin transition temperature. Finally, we summarize our findings in section \ref{summary}. 

\section{Crystal and magnetic structure of \aFeo{} and Computational Methods} \label{methods}

Hematite, \aFeo{}, has the centrosymmetric corundum structure with crystallographic space group R$\bar{3}$c. The magnetic ground state has AFM-ordered spins, aligned along the $z$ easy-axis in a $+--+$ pattern (Fig. \ref{fig:Unit_cells_bandstruc}a). Between the Morin transition at 263 K and the N\'eel temperature of 960K \cite{Morin1951, Dzyaloshinsky1958}, the spins lie in the plane perpendicular to $\hat{z}$ with the same $+--+$ AFM order and a small canting within the plane, giving a weak net spin moment. This is called a canted AFM or weakly FM phase.

Below the Morin transition, the \emph{magnetic} space group is R$\bar{3}$c, which in the absence of SOC gives the nontrivial spin Laue group $^{1}\bar{3}^{2}m = [E \parallel \bar{3}]+[C_2 \parallel \bar{3}m- \bar{3}]$. Here the operations on the left of the double bars act only in spin space and the operations on the right act only in real space. Furthermore, the group $\bar{3}$ is also called a halving subgroup, as it is formed with half of the elements of $^{1}\bar{3}^{2}m$. This halving subgroup $\bar{3}$ contains the six symmetry operations that leave the ordering of the spins unchanged, in this case, the identity operation, space inversion, the rotations about the $z$ axis by $\frac{2\pi}{3}$ and $\frac{4\pi}{3}$, and the combinations of these rotations with space inversion. The coset $\bar{3}m- \bar{3}$ instead contains those operations that reverse the spins, which are the two-fold screw rotations about the axes $[1, 0, 0]$, $[-\frac{1}{2}, \frac{\sqrt{3}}{2}, 0 ]$ and $[-\frac{1}{2}, -\frac{\sqrt{3}}{2}, 0 ]$ with the corresponding translation of half a unit cell length along the $z$ axis, and the combination of these screw rotations with space inversion, where the directions are indicated in Cartesian coordinates. This Laue group allows for $g$-wave spin splitting. Specifically, the $\frac{2\pi}{3}$ and $\frac{4\pi}{3}$ rotations about the $z$ axis enforce a three-fold rotational symmetry in both real and reciprocal space. 

Our first-principles calculations based on density functional theory (DFT) were performed in the plane-wave basis as implemented in the Vienna ab-initio simulation package (\textsc{vasp}) \cite{kresse_efficient_1996, kresse_efficiency_1996}, within the collinear local spin density approximation (LSDA) \cite{LDA_pz} for the easy-axis magnetic phase below the Morin transition, and with non-collinear spins above the transition. In both cases a Hubbard U correction \cite{Liechtenstein_U} was included, with U = 5.5 eV and J = 0.5 eV. As the NRSS occurs in the absence of SOC, this interaction was not included unless stated explicitly. The projector-augmented wave pseudopotentials \cite{blochl_projector_1994} (valence electrons: Fe $3s^2 3p^6 3d^7 4s^1$, O 2s$^2$2p$^4$, datasets Fe\_sv, O) were used, with a kinetic energy cut-off of 600 eV for the wavefunctions in the collinear phase and 800 eV in the canted magnetic phase. Brillouin zone (BZ) integrations were performed using a uniform $\Gamma$-centered $10\times10\times10$ k-point mesh. With these parameters, we
obtained a spin moment on the Fe atoms of 4.01 $\mu_B$ and an electronic band gap of 2.16 eV, close to the experimentally observed values (4.1-4.2 $\mu_B$ \cite{baron_neutron_2005, hillNeutronDiffractionStudy2008} and 2.14-2.2 eV \cite{mochizuki_electrical_1977, gilbert_band-gap_2009}). We used the DFT relaxed crystal structure for \Feo{} that we obtained in \cite{Verbeek_Hidden_2023}, with rhombohedral lattice constants $a = 5.35 \, \mathrm{\AA}, \thickspace \alpha = 55.25^{\circ}$, deviating less than 1.5\% from the experimental values \cite{hill_neutron_2008}. Our lattice vectors in terms of Cartesian coordinates are
\begin{align}
    a_1 &= a \left( -\sin{\frac{\alpha}{2}}, \frac{1}{\sqrt{3}}\sin{\frac{\alpha}{2}}, \sqrt{\frac{1}{3}(4\cos^{2}{\frac{\alpha}{2}}-1 )} \right), \label{eq:vec1} \\
    a_2 &= a \left(0, - \frac{2}{\sqrt{3}}\sin{\frac{\alpha}{2}}, \sqrt{\frac{1}{3}(4\cos^{2}{\frac{\alpha}{2}}-1 )} \right), \label{eq:vec2}  \\
    a_3 &= a \left(\sin{\frac{\alpha}{2}}, \frac{1}{\sqrt{3}}\sin{\frac{\alpha}{2}},\sqrt{\frac{1}{3}(4\cos^{2}{\frac{\alpha}{2}}-1 )} \right). \label{eq:vec3}
\end{align} 

To describe the high-temperature weakly FM phase of \Feo{}, we constrained the direction of the spins on the Fe ions using the constrained moment routine implemented in \textsc{vasp} \cite{ma_constrained_2015}. 
The angular components of the charge and magnetic multipoles were computed from a decomposition of the DFT-calculated charge and magnetic densities into spherical tensors \cite{cricchio_multipoles_2010, granas_theoretical_2012, spaldin_monopole-based_2013}. 
We constrained multipoles by applying a shift in the local potential using the \textsc{multipyles} code \cite{multipyles}, as described in Ref. \cite{schaufelberger_exploring_2023}.

\section{Results and Discussion} \label{results}

\subsection{Multipole analysis} \label{multipole}
 
As stated earlier, the magnetically ordered phases of $\alpha$-Fe$_2$O$_3$ break the global time-reversal symmetry. This implies that there must be a magnetic multipole with ferroic ordering, akin to the ferroic ordering of magnetic dipoles in ferromagnets. In the low-temperature collinear AFM phase, hematite has no net magnetic dipole so a higher-order ferroically-ordered magnetic multipole must be present. The interaction energy $\mathcal{E}_{\rm int, mag}$ of a system with a magnetic field $\Vec{H}(\Vec{r})$ applied to an arbitrary magnetization density $\Vec{\mu}(\Vec{r})$ is given by:
\begin{widetext}
\begin{align}\nonumber \label{eq:magmultipole}
    -\mathcal{E}_{\rm int, mag} = & \underbrace{\Big( \int \mu_{i} \,d\Vec{r} \Big)}_\text{magnetic dipole} H_i|_{\Vec{r}=\Vec{0}} \thickspace + \thickspace
    \underbrace{\Big( \int \mu_{i} r_{j} \,d\Vec{r} \Big)}_{\substack{\text{magnetoelectric} \\ \text{multipole}}}
    \partial_{j} H_{i}|_{\Vec{r}=\Vec{0}} 
    \thickspace + \thickspace \underbrace{\Big( \int \mu_{i} r_{j} r_{k} \,d\Vec{r} \Big)}_\text{magnetic octupole}  \partial_{j} \partial_{k} H_{i}|_{\Vec{r}=\Vec{0}} \\ 
    & \thickspace + \thickspace {\underbrace{\Big( \int \mu_{i} r_{j} r_{k} r_{l} \,d\Vec{r} \Big)}_\text{magnetic hexadecapole}  \partial_{j} \partial_{k} \partial_{l} H_{i}|_{\Vec{r}=\Vec{0}}} \thickspace + \thickspace \underbrace{\Big( \int \mu_{i} r_{j} r_{k} r_{l} r_{m} \,d\Vec{r} \Big)}_\text{magnetic triakontadipole}  \partial_{j} \partial_{k} \partial_{l} \partial_{m} H_{i}|_{\Vec{r}=\Vec{0}} \thickspace + \thickspace ... ,
\end{align}
\end{widetext}
The terms within the parentheses indicate the magnetic multipoles of successive rank, that is the magnetic dipole (1), magnetoelectric multipole (2), magnetic octupole (3), magnetic hexadecapole (4), and magnetic triakontadipole (5) moments respectively (rank as indicated in the parentheses). Note that the second-order magnetic multipole is often called the magnetoelectric multipole due to its association with the linear magnetoelectric effect \cite{spaldin_monopole-based_2013}. As evidenced from Eq. (\ref{eq:magmultipole}), higher-order magnetic multipoles, characterizing the asymmetries and anisotropies in the magnetization density, can still have a ferroic ordering even in the absence of any net magnetic dipole. 

To search for any such ferroically ordered magnetic multipole in hematite, we next perform explicit multipole calculations (see Section \ref{methods} for details). Focusing on the local multipoles centered on the Fe ions, our calculations show that in the absence of SOC, the lowest order magnetic multipole with ferroic ordering is the rank-5 magnetic triakontadipole moment, specifically the $w^{415}_{3}, w^{414}_{-3}$, and $w^{413}_{3}$ irreducible (IR) spherical tensor components (see Table \ref{tab:multipole_components}, block a). Here, in the IR spherical tensor component $w^{k p r}_t$, $k, p, r$ denote respectively the spatial index, spin index (i.e. $p = 0$ for charge and $p = 1$ for magnetic multipoles) and the rank ($r \in \{ | k - p |, | k - p | + 1, \hdots , k + p \}$) of the tensor, while $t \in \{ -r, -r + 1, \hdots , r \}$ labels the component of the tensor \cite{cricchio_itinerant_2009, multipole_decomposition, cricchio_multipoles_2010}. 

We note that the other components of the magnetic triakontadipole, as well as all lower-rank magnetic multipoles, if present, have antiferroic arrangements with opposite signs on different Fe ions and, hence, they can not break the global time-reversal symmetry. This suggests that the magnetic triakontadipole is likely responsible for the NRSS. Interestingly, the inclusion of spin-orbit interaction leads to a ferroic ordering of the rank-3 magnetic octupole components $w^{212}_{0}$ and $w^{213}_{3}$. Since these magnetic octupoles are only induced by SOC, they are not relevant for the NRSS. 

\begin{table}[t]
\caption{The computed spherical IR tensor components $w^{kpr}_{t}$ of the relevant charge and magnetic multipoles in $\alpha$-Fe$_2$O$_3$ without SOC and their ordering pattern on the Fe atoms, corresponding to the numbering indicated in Fig. \ref{fig:Unit_cells_bandstruc}a. In $w^{kpr}_{t}$, $p$ labels the spin. As a consequence $w^{011}_{t}$ give the spin moments, which are opposite in sign to the magnetic dipole moments. The different blocks show multipole moments for both AFM domains (a,b), a different AFM order (the magnetic ground state of \Cro{}) (c), for a high symmetry crystal structure in its magnetic ground state (d) and with multipoles induced (e), the high-temperature phase without (f) and with spin canting (g).}
\centering
\resizebox{\linewidth}{!}{\begin{tabular}{cc|c|c|c}
    \hline
      &System & Multipole & $w^{kpr}_t$ & Sign of multipoles on the Fe sites\\
    \hline\hline
      &                    & Spin moment         & $w^{011}_0$    & $+ - - +$\\ \cline{3-5}
      & Hematite low-T     & Charge Hexadecapole & $w^{404}_3$    & $+ - - +$\\ \cline{3-5}
    a)& magnetic structure & Magnetic            & $w^{415}_3$    & $- - - -$\\
      &                    & Triakontadipole     & $w^{414}_{-3}$ & $- - - -$\\
      &                    &                     & $w^{413}_{3}$  & $- - - -$\\
    \hline
    \hline
      &                    & Spin moment         & $w^{011}_0$    & $- + + -$\\ \cline{3-5}
      &Hematite low-T      & Charge Hexadecapole & $w^{404}_3$    & $+ - - +$\\ \cline{3-5}
    b)&magnetic structure  & Magnetic            & $w^{415}_3$    & $+ + + +$\\
      &AFM domain 2        & Triakontadipole     & $w^{414}_{-3}$ & $+ + + +$\\
      &                    &                     & $w^{413}_{3}$  & $+ + + +$\\
    \hline
    \hline
      &                    & Spin moment         & $w^{011}_0$    & $+ - + -$\\ \cline{3-5}
      &Hematite with       & Charge Hexadecapole & $w^{404}_3$    & $+ - - +$\\ \cline{3-5}
    c)&\Cro{} spin ordering& Magnetic            & $w^{415}_3$    & $0\;0\;0\;0$\\
      &                    & Triakontadipole     & $w^{414}_{-3}$ & $0\;0\;0\;0$\\
      &                    &                     & $w^{413}_{3}$  & $0\;0\;0\;0$\\
    \hline
    \hline
      &                     & Spin moment         & $w^{011}_0$    & $+ - - +$\\\cline{3-5}
      &Hematite high        & Charge Hexadecapole & $w^{404}_{3}$  & $- - - -$\\\cline{3-5}
    d)&symmetry crystal     & Magnetic            & $w^{415}_{3}$  & $- + + -$\\
      & structure           & Triakontadipole     & $w^{414}_{-3}$ & $- + + -$\\
      &                     &                     & $w^{413}_{3}$  & $- + + -$\\
    \hline
    \hline
      &                      & Spin moment         & $w^{011}_0$    & $+ - - +$\\ \cline{3-5}
      &Hematite high symmetry& Charge Hexadecapole & $w^{404}_{-3}$ & $- + + - $\\ \cline{3-5}
    e)&crystal structure     & Magnetic            & $w^{415}_3$    & $- - - -$\\
      & induced hexadecapole & Triakontadipole     & $w^{414}_{-3}$ & $- - - -$\\
      & &                    & $w^{413}_{3}$  & $- - - -$\\
    \hline
    \hline
      &                    & Spin moment         & $w^{011}_{-1}$ & $+ - - +$\\ \cline{3-5}
      &                    & Charge Hexadecapole & $w^{404}_{3}$  & $- + + - $\\ \cline{3-5}
      & Hematite high T    & Magnetic            & $w^{415}_{-2}$ & $- - - -$\\
    f)& magnetic structure & Triakontadipole     & $w^{415}_{-4}$ & $- - - -$\\
      & (no canting)       &                     & $w^{414}_{4}$  & $- - - -$\\
      &                    &                     & $w^{414}_{2}$  & $+ + + +$\\
      &                    &                     & $w^{413}_{-2}$ & $+ + + +$\\
    \hline
    \hline
      &                    & Spin moment         & $w^{011}_{-1}$ & $+ - - +$\\ \cline{4-5}
      &                    &                     & $w^{011}_1$    & $- - - -$\\ \cline{3-5}
      & Hematite high T    & Charge Hexadecapole & $w^{404}_{3}$  & $- + + -$\\ \cline{3-5}
    g)& magnetic structure & Magnetic            & $w^{415}_{-2}$ & $- - - -$\\
      &(with canting)      & Triakontadipole     & $w^{415}_{-4}$ & $- - - -$\\
      &                    &                     & $w^{414}_{4}$  & $- - - -$\\
      &                    &                     & $w^{414}_{2}$  & $+ + + +$\\
      &                    &                     & $w^{413}_{-2}$ & $+ + + +$\\
    \hline
\end{tabular}}
\label{tab:multipole_components}
\end{table}

We point out the correlation between the magnetic triakontadipole and the rank-4 charge hexadecapole $\int r_{i} r_{j} r_{k} r_{l} \,d\Vec{r}$, defined in the expansion of the interaction energy between an arbitrary charge density and an external electric field. By computing the charge multipoles, which quantify the angular distribution of the electronic charge density, we find that the charge hexadecapole component $w^{404}_{3}$ has an antiferroic arrangement on the Fe ions with the same pattern as the $z$ component of the Fe spin moment (see Table \ref{tab:multipole_components}).
This antiferroic pattern of the charge hexadecapoles in combination with the AFM spins gives rise to a ferroic ordering of magnetic triakontadipoles. Note that in the definition of $w^{kpr}_{t}$, $p$ represents the spin, such that $w^{011}_{t}$ gives the spin moments, which are opposite in sign to the magnetic dipole moments. The triakontadipole components $w^{41r}_{t}$ are also given in terms of the spin, rather than the magnetic moments. 

\subsection{Spin splitting in the band structure of hematite} 

Next, we calculate the band structure of hematite in the absence of SOC. As expected from the nontrivial spin Laue group $^{1}\bar{3}^{2}m$, the spin splitting is zero at those points for which the little group contains elements of the coset $\bar{3}m- \bar{3}$ (the two-fold screw rotations). This includes the high symmetry points $\Gamma$, $F$, $L$, $L_1$, $P$, $P_1$, $P_2$, $X$, $Z$ of the rhombohedral BZ, and the high symmetry paths between three sets of these points: [$\Gamma$, $F$, $L_1$, $P_1$, $P_2$, $Z$], [$\Gamma$, $L$, $P$, $Z$] and [$\Gamma$, $X$]. In other regions in k-space, we observe NRSS in all bands. 
 
\begin{figure}[t]
    \centering
    \includegraphics[width = \linewidth]{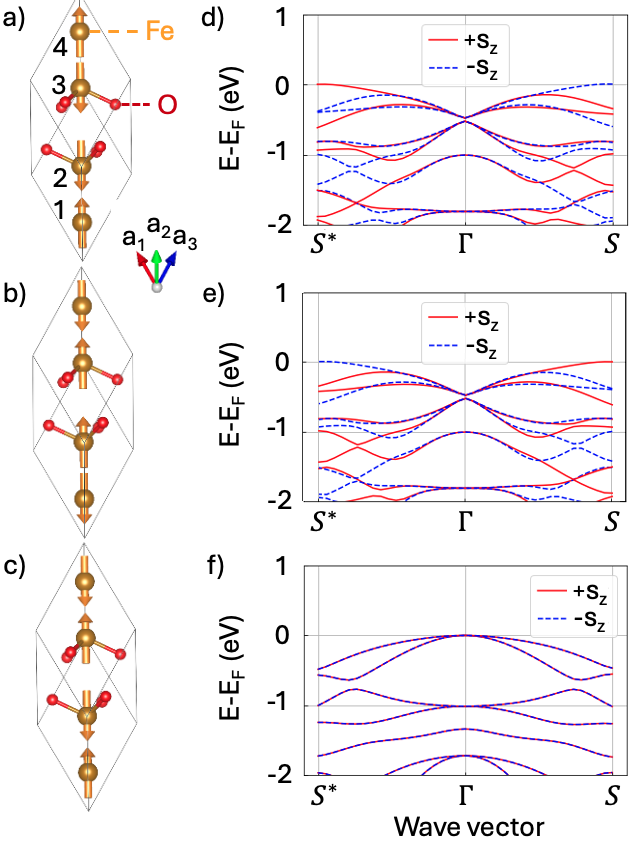}
    \caption{Different configurations of the spins on the Fe atoms and the corresponding spin splitting in \aFeo{}. (a) The primitive unit cell of \aFeo{} and its ground state spin ordering ($+-+-$). (b) The opposite domain with the opposite orientation of the spin moments at the Fe atoms ($-+-+$). (c) A ($+-+-$) spin ordering, the ground state spin ordering of isostructural antiferromagnet \Cro{} (d)-(f) Band structures along the path S$^*$-$\Gamma$-S for the spin ordering shown in (a), (b) and (c) respectively, with $S = (-\frac{1}{12}, \frac{1}{3}, -\frac{1}{4}), \Gamma = (0, 0, 0), S^* = (-0.241, 0.337, -0.096)$ in terms of the reciprocal lattice vectors.} \label{fig:Unit_cells_bandstruc}
\end{figure}

A representative low-symmetry path with large spin splitting is shown in Fig. \ref{fig:Unit_cells_bandstruc}d. Here $S$ and $S^*$ are the points $(-\frac{1}{12}, \frac{1}{3}, -\frac{1}{4})$ and  $(-0.241, 0.337, -0.096)$ in relative coordinates of the reciprocal lattice vectors constructed from the real space vectors defined in Eqs. \ref{eq:vec1}-\ref{eq:vec3}. $S^{*}$ is obtained by rotating $S$ by an angle of $30^\circ$ about the $k_z$ axis. We point out that the NRSS is opposite in sign, but equal in magnitude at $S$ and $S^{*}$. As expected, the spin splitting is zero at $\Gamma$. Note that, while there are additional band crossings along these paths, these crossing points are not dictated by symmetry. \\

\subsection{Manipulation of magnetic triakontadipole and the resulting spin splitting} \label{tuning}

To establish the relation between the magnetic triakontadipoles and the NRSS, we next manipulate the size and sign of the magnetic triakontadipoles and compute the consequent changes in the spin splitting. 

\subsubsection{Change in magnetic ordering}
Perhaps the most straightforward way to change the magnetic triakontadipoles is to change their sign by performing a time reversal operation which can be achieved by flipping the orientation of the spins in our calculation (see Fig. \ref{fig:Unit_cells_bandstruc}b). Physically, this corresponds to the opposite AFM domain. The computed multipoles for this magnetic configuration show that this operation indeed, changes the sign of all magnetic multipoles, while keeping the signs of charge multipoles, including the charge hexadecapole, unchanged (see Table \ref{tab:multipole_components}, block b). We find, as expected, that the spin splitting changes sign as well. Specifically, the $+s_{z}$ and $-s_{z}$ bands are interchanged for the path $S^{*}-\Gamma-S$ (Fig. \ref{fig:Unit_cells_bandstruc}e). 

Next, we manipulate the spin ordering, by changing it to a $+-+-$ pattern (Fig. \ref{fig:Unit_cells_bandstruc}c), the magnetic ground state of the isostructural antiferromagnet \Cro{}\cite{samuelsenInelasticNeutronScattering1970a,brownDeterminationMagnetizationDistribution2002}. We find that this spin ordering leads to an antiferroic pattern of magnetic triakontadipoles, giving rise to a zero net magnetic triakontadipole in the unit cell (see Table \ref{tab:multipole_components}, block c). The corresponding computed band structure has no spin splitting anywhere in the BZ (Fig. \ref{fig:Unit_cells_bandstruc}f), again confirming the correlation between ferroic magnetic triakontadipoles and the NRSS. 

\subsubsection{Change in crystal structure}

\begin{figure}[t]
    \centering
    \includegraphics[width=\linewidth]{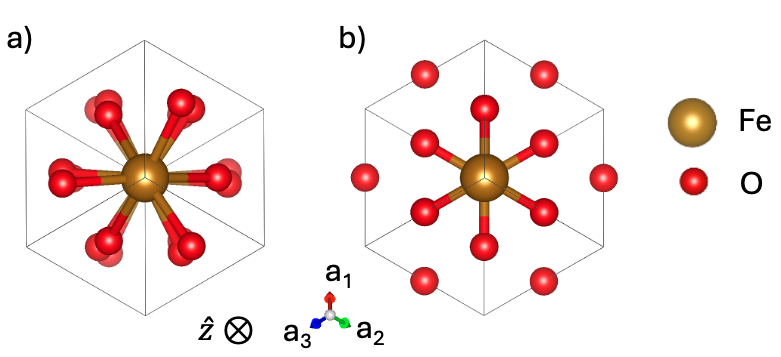}
    \caption{(a) Crystal structure of \Feo{}, and (b) modified high-symmetry crystal structure, both viewed along the $z$ axis. }
    \label{fig:Change_crystal}
\end{figure}

Next, we change the crystal symmetry by shifting the oxygen atoms from their original Wyckoff site symmetry $18e$ to $18d$, while keeping the positions of the Fe atoms fixed (Fig. \ref{fig:Change_crystal}a and b). In this higher-symmetry space group R$\bar{3}$m, ferroically ordered magnetic triakontadipoles are not allowed. 

\begin{figure}[t]
    \centering
    \includegraphics[width=\linewidth]{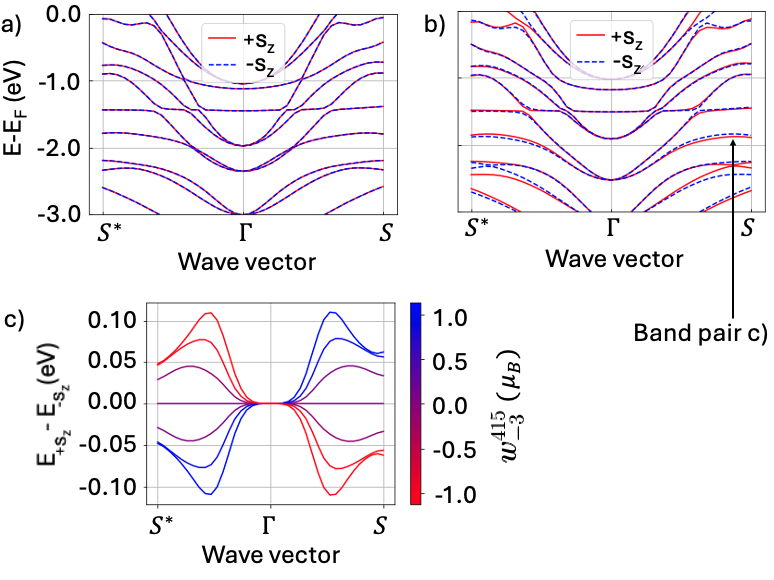}
    \caption{Band structures for the R$\bar{3}$m structure without induced ferroically magnetic triakontadipoles (a) and with induced magnetic triakontadipole component $w^{415}_{-3} = -0.087 \thickspace \mu_B$ (b). (c) shows the variation in the induced spin splitting between a specific pair of bands on manipulating the size of the angular component $w^{415}_{-3}$ of the magnetic triakontadipole (color bar).  The zero value of $w^{415}_{-3}$ corresponds to the relaxed electronic structure with no constrained multipole moments. The pair of bands used in (c) is indicated with a black arrow in (b).}
    \label{fig:BS_Opos_shift}
\end{figure}

We perform DFT calculations with the R$\bar{3}$m structure and the original $+-+-$ ordering of the spin on the Fe atoms, and find antiferroically ordered magnetic triakontadipoles (see Table \ref{tab:multipole_components}, block d ). This is the result of ferroically ordered charge hexadecapoles combined with the antiferroic ordering of the spins. 
Consequently, we find no NRSS (Fig. \ref{fig:BS_Opos_shift}a ). Our results highlight the importance of the surrounding nonmagnetic environment in establishing and manipulating both the magnetic triakontadipole and the $g$-wave NRSS in hematite. Similarly, the influence of the nonmagnetic environment on $d$-wave NRSS has also been emphasized \cite{Libor2020, BhowalSpaldin2024}.

\subsubsection{Constraining multipoles} 

We now use the constrained multipole method mentioned in Section \ref{methods} to manipulate the size of the ferroically ordered magnetic triakontadipoles in the actual $R\bar{3}c$ structure of hematite, and also to introduce them into the hypothetical $R\bar{3}m$ structure. We achieve this by manipulating the charge multipole of one order lower, i.e. the charge hexadecapoles. 

\begin{figure}[h]
    \centering
    \includegraphics[width=\linewidth]{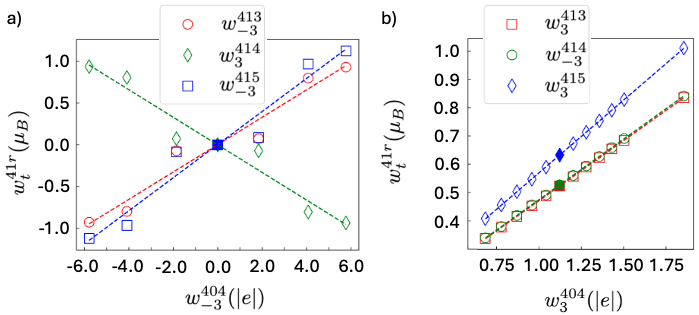}
    \caption{The magnitude of the angular component of the ferroically ordered magnetic triakontadipole components (in units of $\mu_B$) as a function of the magnitude of the angular component of the constrained antiferroically ordered charge hexadecapole component (in units of $|e|$ ). a) the hypothetical R$\bar{3}$m structure, with the charge hexadecapole component $w^{404}_{-3}$ constrained, and b) the actual \aFeo{} structure with the $w^{404}_{3}$ hexadecapole component constrained. The dashed lines indicate the linear fit as a guide to the eye. The filled markers indicate the values of the multipoles without any constraints.}
    \label{fig:Multipole_dependence}
\end{figure}

We show the relation between the magnitude of the ferroically ordered triakontadipoles and the size of the constrained hexadecapoles in both structures in Fig. \ref{fig:Multipole_dependence}. Note that $w^{kpr}_{t}$ tensors capture only the angular part of the multipole, i.e. without performing the radial part of the integration in Eq. \ref{eq:magmultipole}. Thus, the magnitude of the components is given is $\mu_B$ for the magnetic triakontadipoles and in terms of the electronic charge $|e|$ for the charge hexadecapoles. We see that in both structures the size of the magnetic triakontadipoles increases proportionally with the increase in the charge hexadecapole, with an approximate linear dependence, particularly over small ranges. However, there are some deviations from this trend in the hypothetical high-symmetry structure. We note that we induce a different antiferroically ordered charge hexadecapole component in the hypothetical high-symmetry structure compared to the actual \aFeo{} structure (since the same one already exists with ferroic ordering), so the introduced ferroically ordered magnetic triakontadipole components are also different. 

We now discuss the consequences of constraining these multipoles. First, by constraining the charge hexadecapole component $w^{404}_{-3}$ to be non-zero and to have the same antiferroic ordering as the spins in the hypothetical $R\bar{3}m$ structure, we induce the ferroically ordered magnetic triakontadipoles $w^{415}_{3}$, $w^{414}_{-3}$ and $w^{413}_{3}$ (Table \ref{tab:multipole_components}, block e). The computed bands show the presence of NRSS (Fig. \ref{fig:BS_Opos_shift}b), with equal and opposite splitting along $S^{*}-\Gamma$ and $\Gamma-S$, as expected. 

Next, we vary the sign and magnitude of the induced charge hexadecapole while keeping the antiferroic ordering pattern the same. In Fig. \ref{fig:BS_Opos_shift}c, we see that with increasing magnitude of the magnetic triakontadipole $w^{415}_{-3}$, the magnitude of the NRSS increases, and when $w^{415}_{-3}$ switches sign, so does the NRSS. This further confirms that the NRSS is driven by the magnetic triakontadipoles. 

\begin{figure}[t]
    \centering
    \includegraphics[width=\linewidth]{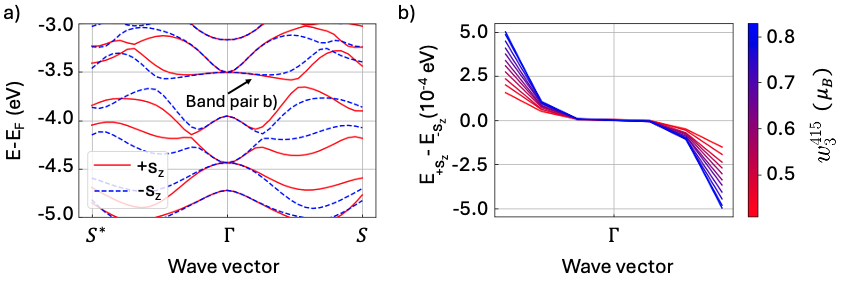}
    \caption{(a) Spin-polarized band structure of hematite, with an arrow indicating a pair of bands. (b) Spin splitting near $\Gamma$, for the bands indicated in (a) with a black arrow (approximately $-3.5$ eV below the Fermi energy at $\Gamma$), plotted as a function of the magnitude (in $\mu_B$) of $w^{415}_{3}$, the angular component of one of the ferroically ordered magnetic triakontadipole components.}
    \label{fig:BS_Fe_shift_gamma_}
\end{figure}

Furthermore, we show the effect of constraining the multipoles in the actual R$\bar{3}$c structure on the NRSS, analyzing the spin-split bands in the vicinity of the $\Gamma$ point, where the bands, which have many crossings, can be easily distinguished. As expected, we see in Fig. \ref{fig:BS_Fe_shift_gamma_}, that the NRSS increases with the increasing value of magnetic triakontadipole, further confirming the magnetic triakontadipole to be responsible for the spin splitting in hematite. We find the same behavior at smaller values of the Hubbard U, where the bands are less entangled and the change in NRSS is easier to distinguish.

\subsection{Weakly FM phase of $\alpha$-Fe$_2$O$_3$}

Finally, as discussed in Sec. \ref{methods}, above the 263 K Morin transition, the spins in hematite lie in the $x$-$y$ plane, oriented in a $+--+$ AFM pattern along the $y$ axis, with small ferroically ordered components along $x$ \cite{Dzyaloshinsky1958}. This spin ordering has a lower symmetry than the ground state. We explore the multipoles and the corresponding NRSS in this high-temperature magnetic phase, in two steps. First, we orient the spins along the $y$ axis and constrain them to be collinear and antiferromagnetically ordered. With this spin ordering, we find several ferroically ordered triakontadipoles (Tab. \ref{tab:multipole_components}, block f), with different components than in the low-temperature ground state, and a NRSS with the bands now split in spin parallel or antiparallel to $\hat{y}$ (e.g. $+s_{y}$ and $-s_{y}$). 

\begin{figure}[t]
    \centering
    \includegraphics[width=\linewidth]{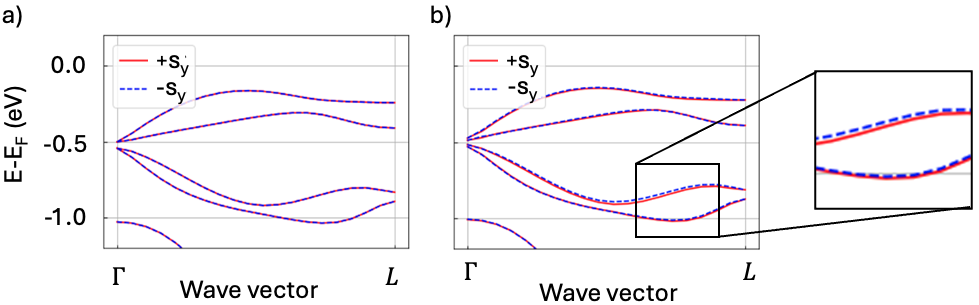}
    \caption{Band structure of $\alpha$-Fe$_2$O$_3$ along a high symmetry path where splitting due to the ferroically ordered triakontadipoles is forbidden, (a) for an AFM ordering with the moments oriented along the $\hat{y}$ in a $+--+$ pattern and (b) for the high-temperature magnetic phase with canted moment. The inset shows a zoomed-in view of the splitting between spin-polarized bands in (b). The bands are shown along the path $\Gamma \rightarrow L$, where $L\equiv (\frac{1}{2}, 0, 0)$ in relative coordinates, positioned at the center of the hexagonal face of the Brillouin Zone (see Fig. \ref{fig:Unit_cells_bandstruc}c).} 
    \label{fig:BS_cant_nc}
\end{figure}

Next, we constrain the spins to allow for a small FM component along $\hat{x}$ of $\sim$0.1 $\mu_B$ per Fe atom. We note that this is larger than the experimentally observed $10^{-3} \thickspace \mu_B$ moment in hematite \cite{flanders_magnetic_1965, bodker_magnetic_2000}. The resulting multipoles are shown in Table \ref{tab:multipole_components}, block g. In addition to the same ferroically ordered triakontadipole components as in the collinear AFM along $\hat{y}$ case. We now have a ferroic ordering of spins parallel to $\hat{x}$, and nine further small ferroically ordered triakontadipole components (not listed in Tab. \ref{tab:multipole_components}). Now, we look at the NRSS in the weakly FM phase and compare it to the situation where the spins are collinear AFM along $y$. We observe that in the weak FM phase, there is a small additional splitting due to the ferroic ordering of the spins. This additional splitting can best be seen along directions in which the NRSS due to the ferroic ordering of the triakontadipoles is absent, such as the $\Gamma-L$ direction. We see an absence of spin splitting for the collinear AFM arrangement, with moment along $\hat{y}$ (Fig. \ref{fig:BS_cant_nc}a), and a small splitting upon canting (Fig. \ref{fig:BS_cant_nc}b), induced by the weak ferromagnetism.

\section{Summary and Outlook \label{summary}} 

In summary, we have identified a ferroically ordered rank-5 magnetic triakontadipole in the low-temperature AFM phase of hematite. The ferroically ordered magnetic triakontadipole is present even in the absence of SOC, where it is the lowest-order time-reversal symmetry breaking multipole. As a result, it causes an NRSS with the same $g$-wave symmetry as the magnetic triakontadipoles. The ferroically ordered magnetic triakontadipoles persist in the weakly ferromagnetic phase above the Morin transition, where they dominate over the ferroically aligned spin components in their contribution to the spin splitting. Our findings indicate a method to manipulate both the sign and magnitude of the NRSS by altering the crystal structure and magnetic ordering. As such, they contribute to the recent efforts \cite{vanderLaan2023,Lee2024,Krempask2024,reimers_direct_2024} aimed at harnessing $g$-wave spin splitting with specific implications for the unconventional transport properties observed in $\alpha$-Fe$_2$O$_3$ \cite{Fischer2020,Kanj2023,Lebrun2020}.
The manifestation of the magnetic triakontadipole in the NRSS of hematite may be probed using spin-polarized angle-resolved photoemission spectroscopy, as has recently been reported to demonstrate the spin splitting in other AFM materials \cite{Lee2024,Krempask2024,reimers_direct_2024,Lin2024}. The presence of the magnetic triakontadipoles and the NRSS may also be the source of some unexplained features in the spin-wave spectra of \aFeo{} \cite{Kanj2023}. 
\\
We emphasize that the relevance of the ferroically-ordered magnetic triakontadipoles goes beyond the NRSS. For example, they allow for a fourth-order magnetoelectric effect, where the induced magnetisation is quartic in the applied electric field. Furthermore, the magnetic triakontadipoles give rise to a second-order piezomagnetic effect, where the induced magnetisation scales quadratically with the applied strain \cite{mcclarty_landau_2024, aoyama_piezomagnetic_2023}. Similarly, the SOC-induced magnetic octupole in hematite, as discussed in the present work, gives rise to the second-order magnetoelectric effect \cite{Verbeek_Hidden_2023, Urru2022} and piezomagnetic effect \cite{BhowalSpaldin2024}, which may also be probed experimentally. 
\\
Finally, our present understanding implies the rank-7 inversion symmetric magnetic multipole will be responsible for the $i$-wave spin splitting, which requires future investigation. We hope that our work will stimulate further theoretical and experimental work along these directions.

\begin{acknowledgments}
The authors thank Dr. Andrea Urru for useful discussions. NAS, XHV, and SB were supported by the ERC under the European Union’s Horizon 2020 research and innovation programme grant No. 810451 and by the ETH Z\"urich. Computational resources were provided by ETH Z\"urich’s Euler cluster.
\end{acknowledgments}

\vfill

\bibliography{references,sb}

\providecommand{\noopsort}[1]{}\providecommand{\singleletter}[1]{#1}
\begin{thebibliography}{64}%
\makeatletter
\providecommand \@ifxundefined [1]{%
 \@ifx{#1\undefined}
}%
\providecommand \@ifnum [1]{%
 \ifnum #1\expandafter \@firstoftwo
 \else \expandafter \@secondoftwo
 \fi
}%
\providecommand \@ifx [1]{%
 \ifx #1\expandafter \@firstoftwo
 \else \expandafter \@secondoftwo
 \fi
}%
\providecommand \natexlab [1]{#1}%
\providecommand \enquote  [1]{``#1''}%
\providecommand \bibnamefont  [1]{#1}%
\providecommand \bibfnamefont [1]{#1}%
\providecommand \citenamefont [1]{#1}%
\providecommand \href@noop [0]{\@secondoftwo}%
\providecommand \href [0]{\begingroup \@sanitize@url \@href}%
\providecommand \@href[1]{\@@startlink{#1}\@@href}%
\providecommand \@@href[1]{\endgroup#1\@@endlink}%
\providecommand \@sanitize@url [0]{\catcode `\\12\catcode `\$12\catcode `\&12\catcode `\#12\catcode `\^12\catcode `\_12\catcode `\%12\relax}%
\providecommand \@@startlink[1]{}%
\providecommand \@@endlink[0]{}%
\providecommand \url  [0]{\begingroup\@sanitize@url \@url }%
\providecommand \@url [1]{\endgroup\@href {#1}{\urlprefix }}%
\providecommand \urlprefix  [0]{URL }%
\providecommand \Eprint [0]{\href }%
\providecommand \doibase [0]{https://doi.org/}%
\providecommand \selectlanguage [0]{\@gobble}%
\providecommand \bibinfo  [0]{\@secondoftwo}%
\providecommand \bibfield  [0]{\@secondoftwo}%
\providecommand \translation [1]{[#1]}%
\providecommand \BibitemOpen [0]{}%
\providecommand \bibitemStop [0]{}%
\providecommand \bibitemNoStop [0]{.\EOS\space}%
\providecommand \EOS [0]{\spacefactor3000\relax}%
\providecommand \BibitemShut  [1]{\csname bibitem#1\endcsname}%
\let\auto@bib@innerbib\@empty
\bibitem [{\citenamefont {Pekar}\ and\ \citenamefont {Rashba}(1964)}]{PekarRashba1964}%
  \BibitemOpen
  \bibfield  {author} {\bibinfo {author} {\bibfnamefont {S.}~\bibnamefont {Pekar}}\ and\ \bibinfo {author} {\bibfnamefont {{\'E}.~I.}\ \bibnamefont {Rashba}},\ }\bibfield  {title} {\bibinfo {title} {Combined resonance in crystals in inhomogeneous magnetic fields},\ }\href {http://www.jetp.ras.ru/cgi-bin/e/index/r/47/5/p1927?a=list} {\bibfield  {journal} {\bibinfo  {journal} {Zh. Eksp. Teor. Fiz.}\ }\textbf {\bibinfo {volume} {47}},\ \bibinfo {pages} {1927} (\bibinfo {year} {1964})},\ \bibinfo {note} {[Sov. Phys. JETP 20 (5), 1295-1298 (1965)]}\BibitemShut {NoStop}%
\bibitem [{\citenamefont {Rashba}(1960)}]{Rashba1960}%
  \BibitemOpen
  \bibfield  {author} {\bibinfo {author} {\bibfnamefont {E.}~\bibnamefont {Rashba}},\ }\bibfield  {title} {\bibinfo {title} {Properties of semiconductors with an extremum loop. i. cyclotron and combinational resonance in a magnetic field perpendicular to the plane of the loop},\ }\href {https://cir.nii.ac.jp/crid/1571698600346713472} {\bibfield  {journal} {\bibinfo  {journal} {Sov. Phys.-Solid State}\ }\textbf {\bibinfo {volume} {2}},\ \bibinfo {pages} {1109} (\bibinfo {year} {1960})}\BibitemShut {NoStop}%
\bibitem [{\citenamefont {Bychkov}\ and\ \citenamefont {Rashba}(1984)}]{BychkovRashba}%
  \BibitemOpen
  \bibfield  {author} {\bibinfo {author} {\bibfnamefont {Y.~A.}\ \bibnamefont {Bychkov}}\ and\ \bibinfo {author} {\bibfnamefont {E.~I.}\ \bibnamefont {Rashba}},\ }\bibfield  {title} {\bibinfo {title} {Oscillatory effects and the magnetic susceptibility of carriers in inversion layers},\ }\href {https://doi.org/10.1088/0022-3719/17/33/015} {\bibfield  {journal} {\bibinfo  {journal} {J. Solid State Phys.}\ }\textbf {\bibinfo {volume} {17}},\ \bibinfo {pages} {6039} (\bibinfo {year} {1984})}\BibitemShut {NoStop}%
\bibitem [{\citenamefont {Šmejkal}\ \emph {et~al.}(2020)\citenamefont {Šmejkal}, \citenamefont {González-Hernández}, \citenamefont {Jungwirth},\ and\ \citenamefont {Sinova}}]{Libor2020}%
  \BibitemOpen
  \bibfield  {author} {\bibinfo {author} {\bibfnamefont {L.}~\bibnamefont {Šmejkal}}, \bibinfo {author} {\bibfnamefont {R.}~\bibnamefont {González-Hernández}}, \bibinfo {author} {\bibfnamefont {T.}~\bibnamefont {Jungwirth}},\ and\ \bibinfo {author} {\bibfnamefont {J.}~\bibnamefont {Sinova}},\ }\bibfield  {title} {\bibinfo {title} {Crystal time-reversal symmetry breaking and spontaneous {H}all effect in collinear antiferromagnets},\ }\href {https://doi.org/10.1126/sciadv.aaz8809} {\bibfield  {journal} {\bibinfo  {journal} {Sci. Adv.}\ }\textbf {\bibinfo {volume} {6}},\ \bibinfo {pages} {eaaz8809} (\bibinfo {year} {2020})}\BibitemShut {NoStop}%
\bibitem [{\citenamefont {Naka}\ \emph {et~al.}(2019)\citenamefont {Naka}, \citenamefont {Hayami}, \citenamefont {Kusunose}, \citenamefont {Yanagi}, \citenamefont {Motome},\ and\ \citenamefont {Seo}}]{Naka2019}%
  \BibitemOpen
  \bibfield  {author} {\bibinfo {author} {\bibfnamefont {M.}~\bibnamefont {Naka}}, \bibinfo {author} {\bibfnamefont {S.}~\bibnamefont {Hayami}}, \bibinfo {author} {\bibfnamefont {H.}~\bibnamefont {Kusunose}}, \bibinfo {author} {\bibfnamefont {Y.}~\bibnamefont {Yanagi}}, \bibinfo {author} {\bibfnamefont {Y.}~\bibnamefont {Motome}},\ and\ \bibinfo {author} {\bibfnamefont {H.}~\bibnamefont {Seo}},\ }\bibfield  {title} {\bibinfo {title} {Spin current generation in organic antiferromagnets},\ }\href {https://doi.org/10.1038/s41467-019-12229-y} {\bibfield  {journal} {\bibinfo  {journal} {Nat. Commun.}\ }\textbf {\bibinfo {volume} {10}},\ \bibinfo {pages} {4305} (\bibinfo {year} {2019})}\BibitemShut {NoStop}%
\bibitem [{\citenamefont {Ahn}\ \emph {et~al.}(2019)\citenamefont {Ahn}, \citenamefont {Hariki}, \citenamefont {Lee},\ and\ \citenamefont {Kune\ifmmode~\check{s}\else \v{s}\fi{}}}]{Kyo-Hoon2019}%
  \BibitemOpen
  \bibfield  {author} {\bibinfo {author} {\bibfnamefont {K.-H.}\ \bibnamefont {Ahn}}, \bibinfo {author} {\bibfnamefont {A.}~\bibnamefont {Hariki}}, \bibinfo {author} {\bibfnamefont {K.-W.}\ \bibnamefont {Lee}},\ and\ \bibinfo {author} {\bibfnamefont {J.}~\bibnamefont {Kune\ifmmode~\check{s}\else \v{s}\fi{}}},\ }\bibfield  {title} {\bibinfo {title} {Antiferromagnetism in {RuO}$_{2}$ as $d$-wave {Pomeranchuk} instability},\ }\href {https://doi.org/10.1103/PhysRevB.99.184432} {\bibfield  {journal} {\bibinfo  {journal} {Phys. Rev. B}\ }\textbf {\bibinfo {volume} {99}},\ \bibinfo {pages} {184432} (\bibinfo {year} {2019})}\BibitemShut {NoStop}%
\bibitem [{\citenamefont {Hayami}\ \emph {et~al.}(2019)\citenamefont {Hayami}, \citenamefont {Yanagi},\ and\ \citenamefont {Kusunose}}]{Hayami2019}%
  \BibitemOpen
  \bibfield  {author} {\bibinfo {author} {\bibfnamefont {S.}~\bibnamefont {Hayami}}, \bibinfo {author} {\bibfnamefont {Y.}~\bibnamefont {Yanagi}},\ and\ \bibinfo {author} {\bibfnamefont {H.}~\bibnamefont {Kusunose}},\ }\bibfield  {title} {\bibinfo {title} {Momentum-dependent spin splitting by collinear antiferromagnetic ordering},\ }\href {https://doi.org/10.7566/JPSJ.88.123702} {\bibfield  {journal} {\bibinfo  {journal} {J. Phys. Soc. Jpn.}\ }\textbf {\bibinfo {volume} {88}},\ \bibinfo {pages} {123702} (\bibinfo {year} {2019})}\BibitemShut {NoStop}%
\bibitem [{\citenamefont {Yuan}\ \emph {et~al.}(2020)\citenamefont {Yuan}, \citenamefont {Wang}, \citenamefont {Luo}, \citenamefont {Rashba},\ and\ \citenamefont {Zunger}}]{Yuan2020}%
  \BibitemOpen
  \bibfield  {author} {\bibinfo {author} {\bibfnamefont {L.-D.}\ \bibnamefont {Yuan}}, \bibinfo {author} {\bibfnamefont {Z.}~\bibnamefont {Wang}}, \bibinfo {author} {\bibfnamefont {J.-W.}\ \bibnamefont {Luo}}, \bibinfo {author} {\bibfnamefont {E.~I.}\ \bibnamefont {Rashba}},\ and\ \bibinfo {author} {\bibfnamefont {A.}~\bibnamefont {Zunger}},\ }\bibfield  {title} {\bibinfo {title} {Giant momentum-dependent spin splitting in centrosymmetric low-$z$ antiferromagnets},\ }\href {https://doi.org/10.1103/PhysRevB.102.014422} {\bibfield  {journal} {\bibinfo  {journal} {Phys. Rev. B}\ }\textbf {\bibinfo {volume} {102}},\ \bibinfo {pages} {014422} (\bibinfo {year} {2020})}\BibitemShut {NoStop}%
\bibitem [{\citenamefont {Yuan}\ \emph {et~al.}(2021)\citenamefont {Yuan}, \citenamefont {Wang}, \citenamefont {Luo},\ and\ \citenamefont {Zunger}}]{Yuan2021}%
  \BibitemOpen
  \bibfield  {author} {\bibinfo {author} {\bibfnamefont {L.-D.}\ \bibnamefont {Yuan}}, \bibinfo {author} {\bibfnamefont {Z.}~\bibnamefont {Wang}}, \bibinfo {author} {\bibfnamefont {J.-W.}\ \bibnamefont {Luo}},\ and\ \bibinfo {author} {\bibfnamefont {A.}~\bibnamefont {Zunger}},\ }\bibfield  {title} {\bibinfo {title} {Prediction of low-z collinear and noncollinear antiferromagnetic compounds having momentum-dependent spin splitting even without spin-orbit coupling},\ }\href {https://doi.org/10.1103/PhysRevMaterials.5.014409} {\bibfield  {journal} {\bibinfo  {journal} {Phys. Rev. Materials}\ }\textbf {\bibinfo {volume} {5}},\ \bibinfo {pages} {014409} (\bibinfo {year} {2021})}\BibitemShut {NoStop}%
\bibitem [{\citenamefont {Šmejkal}\ \emph {et~al.}(2022)\citenamefont {Šmejkal}, \citenamefont {Sinova},\ and\ \citenamefont {Jungwirth}}]{Smejkal2022PRX}%
  \BibitemOpen
  \bibfield  {author} {\bibinfo {author} {\bibfnamefont {L.}~\bibnamefont {Šmejkal}}, \bibinfo {author} {\bibfnamefont {J.}~\bibnamefont {Sinova}},\ and\ \bibinfo {author} {\bibfnamefont {T.}~\bibnamefont {Jungwirth}},\ }\bibfield  {title} {\bibinfo {title} {Beyond conventional ferromagnetism and antiferromagnetism: A phase with nonrelativistic spin and crystal rotation symmetry},\ }\href {https://doi.org/10.1103/PhysRevX.12.031042} {\bibfield  {journal} {\bibinfo  {journal} {Phys. Rev. X}\ }\textbf {\bibinfo {volume} {12}},\ \bibinfo {pages} {031042} (\bibinfo {year} {2022})}\BibitemShut {NoStop}%
\bibitem [{\citenamefont {Yuan}\ and\ \citenamefont {Zunger}(2023)}]{yuan_degeneracy_2023}%
  \BibitemOpen
  \bibfield  {author} {\bibinfo {author} {\bibfnamefont {L.-D.}\ \bibnamefont {Yuan}}\ and\ \bibinfo {author} {\bibfnamefont {A.}~\bibnamefont {Zunger}},\ }\bibfield  {title} {\bibinfo {title} {Degeneracy {Removal} of {Spin} {Bands} in {Collinear} {Antiferromagnets} with {Non}-{Interconvertible} {Spin}-{Structure} {Motif} {Pair}},\ }\href {https://doi.org/10.1002/adma.202211966} {\bibfield  {journal} {\bibinfo  {journal} {Adv Mater}\ }\textbf {\bibinfo {volume} {35}},\ \bibinfo {pages} {e2211966} (\bibinfo {year} {2023})}\BibitemShut {NoStop}%
\bibitem [{\citenamefont {Yuan}\ \emph {et~al.}(2023)\citenamefont {Yuan}, \citenamefont {Zhang}, \citenamefont {Acosta},\ and\ \citenamefont {Zunger}}]{yuan_uncovering_2023}%
  \BibitemOpen
  \bibfield  {author} {\bibinfo {author} {\bibfnamefont {L.-D.}\ \bibnamefont {Yuan}}, \bibinfo {author} {\bibfnamefont {X.}~\bibnamefont {Zhang}}, \bibinfo {author} {\bibfnamefont {C.~M.}\ \bibnamefont {Acosta}},\ and\ \bibinfo {author} {\bibfnamefont {A.}~\bibnamefont {Zunger}},\ }\bibfield  {title} {\bibinfo {title} {Uncovering spin-orbit coupling-independent hidden spin polarization of energy bands in antiferromagnets},\ }\href {https://doi.org/10.1038/s41467-023-40877-8} {\bibfield  {journal} {\bibinfo  {journal} {Nat Commun}\ }\textbf {\bibinfo {volume} {14}},\ \bibinfo {pages} {5301} (\bibinfo {year} {2023})}\BibitemShut {NoStop}%
\bibitem [{\citenamefont {\ifmmode~\check{S}\else \v{S}\fi{}mejkal}\ \emph {et~al.}(2022{\natexlab{a}})\citenamefont {\ifmmode~\check{S}\else \v{S}\fi{}mejkal}, \citenamefont {Sinova},\ and\ \citenamefont {Jungwirth}}]{Smejkal2022}%
  \BibitemOpen
  \bibfield  {author} {\bibinfo {author} {\bibfnamefont {L.}~\bibnamefont {\ifmmode~\check{S}\else \v{S}\fi{}mejkal}}, \bibinfo {author} {\bibfnamefont {J.}~\bibnamefont {Sinova}},\ and\ \bibinfo {author} {\bibfnamefont {T.}~\bibnamefont {Jungwirth}},\ }\bibfield  {title} {\bibinfo {title} {Emerging research landscape of altermagnetism},\ }\href {https://doi.org/10.1103/PhysRevX.12.040501} {\bibfield  {journal} {\bibinfo  {journal} {Phys. Rev. X}\ }\textbf {\bibinfo {volume} {12}},\ \bibinfo {pages} {040501} (\bibinfo {year} {2022}{\natexlab{a}})}\BibitemShut {NoStop}%
\bibitem [{\citenamefont {Gonz\'alez-Hern\'andez}\ \emph {et~al.}(2021)\citenamefont {Gonz\'alez-Hern\'andez}, \citenamefont {\ifmmode~\check{S}\else \v{S}\fi{}mejkal}, \citenamefont {V\'yborn\'y}, \citenamefont {Yahagi}, \citenamefont {Sinova}, \citenamefont {Jungwirth},\ and\ \citenamefont {\ifmmode~\check{Z}\else \v{Z}\fi{}elezn\'y}}]{Hernandez2021}%
  \BibitemOpen
  \bibfield  {author} {\bibinfo {author} {\bibfnamefont {R.}~\bibnamefont {Gonz\'alez-Hern\'andez}}, \bibinfo {author} {\bibfnamefont {L.}~\bibnamefont {\ifmmode~\check{S}\else \v{S}\fi{}mejkal}}, \bibinfo {author} {\bibfnamefont {K.}~\bibnamefont {V\'yborn\'y}}, \bibinfo {author} {\bibfnamefont {Y.}~\bibnamefont {Yahagi}}, \bibinfo {author} {\bibfnamefont {J.}~\bibnamefont {Sinova}}, \bibinfo {author} {\bibfnamefont {T.}~\bibnamefont {Jungwirth}},\ and\ \bibinfo {author} {\bibfnamefont {J.}~\bibnamefont {\ifmmode~\check{Z}\else \v{Z}\fi{}elezn\'y}},\ }\bibfield  {title} {\bibinfo {title} {Efficient electrical spin splitter based on nonrelativistic collinear antiferromagnetism},\ }\href {https://doi.org/10.1103/PhysRevLett.126.127701} {\bibfield  {journal} {\bibinfo  {journal} {Phys. Rev. Lett.}\ }\textbf {\bibinfo {volume} {126}},\ \bibinfo {pages} {127701} (\bibinfo {year} {2021})}\BibitemShut {NoStop}%
\bibitem [{\citenamefont {Shao}\ \emph {et~al.}(2021)\citenamefont {Shao}, \citenamefont {Zhang}, \citenamefont {Li}, \citenamefont {Eom},\ and\ \citenamefont {Tsymbal}}]{Shao2021}%
  \BibitemOpen
  \bibfield  {author} {\bibinfo {author} {\bibfnamefont {D.-F.}\ \bibnamefont {Shao}}, \bibinfo {author} {\bibfnamefont {S.-H.}\ \bibnamefont {Zhang}}, \bibinfo {author} {\bibfnamefont {M.}~\bibnamefont {Li}}, \bibinfo {author} {\bibfnamefont {C.-B.}\ \bibnamefont {Eom}},\ and\ \bibinfo {author} {\bibfnamefont {E.~Y.}\ \bibnamefont {Tsymbal}},\ }\bibfield  {title} {\bibinfo {title} {Spin-neutral currents for spintronics},\ }\href {https://doi.org/10.1038/s41467-021-26915-3} {\bibfield  {journal} {\bibinfo  {journal} {Nat. Commun.}\ }\textbf {\bibinfo {volume} {12}},\ \bibinfo {pages} {7061} (\bibinfo {year} {2021})}\BibitemShut {NoStop}%
\bibitem [{\citenamefont {Bose}\ \emph {et~al.}(2022)\citenamefont {Bose}, \citenamefont {Schreiber}, \citenamefont {Jain}, \citenamefont {Shao}, \citenamefont {Nair}, \citenamefont {Sun}, \citenamefont {Zhang}, \citenamefont {Muller}, \citenamefont {Tsymbal}, \citenamefont {Schlom},\ and\ \citenamefont {Ralph}}]{Bose2022}%
  \BibitemOpen
  \bibfield  {author} {\bibinfo {author} {\bibfnamefont {A.}~\bibnamefont {Bose}}, \bibinfo {author} {\bibfnamefont {N.~J.}\ \bibnamefont {Schreiber}}, \bibinfo {author} {\bibfnamefont {R.}~\bibnamefont {Jain}}, \bibinfo {author} {\bibfnamefont {D.-F.}\ \bibnamefont {Shao}}, \bibinfo {author} {\bibfnamefont {H.~P.}\ \bibnamefont {Nair}}, \bibinfo {author} {\bibfnamefont {J.}~\bibnamefont {Sun}}, \bibinfo {author} {\bibfnamefont {X.~S.}\ \bibnamefont {Zhang}}, \bibinfo {author} {\bibfnamefont {D.~A.}\ \bibnamefont {Muller}}, \bibinfo {author} {\bibfnamefont {E.~Y.}\ \bibnamefont {Tsymbal}}, \bibinfo {author} {\bibfnamefont {D.~G.}\ \bibnamefont {Schlom}},\ and\ \bibinfo {author} {\bibfnamefont {D.~C.}\ \bibnamefont {Ralph}},\ }\bibfield  {title} {\bibinfo {title} {Tilted spin current generated by the collinear antiferromagnet ruthenium dioxide},\ }\href {https://doi.org/10.1038/s41928-022-00744-8} {\bibfield  {journal} {\bibinfo  {journal} {Nat. Electron.}\ }\textbf {\bibinfo {volume} {5}},\ \bibinfo {pages}
  {267} (\bibinfo {year} {2022})}\BibitemShut {NoStop}%
\bibitem [{\citenamefont {Bai}\ \emph {et~al.}(2022)\citenamefont {Bai}, \citenamefont {Han}, \citenamefont {Feng}, \citenamefont {Zhou}, \citenamefont {Su}, \citenamefont {Wang}, \citenamefont {Liao}, \citenamefont {Zhu}, \citenamefont {Chen}, \citenamefont {Pan}, \citenamefont {Fan},\ and\ \citenamefont {Song}}]{Bai2022}%
  \BibitemOpen
  \bibfield  {author} {\bibinfo {author} {\bibfnamefont {H.}~\bibnamefont {Bai}}, \bibinfo {author} {\bibfnamefont {L.}~\bibnamefont {Han}}, \bibinfo {author} {\bibfnamefont {X.~Y.}\ \bibnamefont {Feng}}, \bibinfo {author} {\bibfnamefont {Y.~J.}\ \bibnamefont {Zhou}}, \bibinfo {author} {\bibfnamefont {R.~X.}\ \bibnamefont {Su}}, \bibinfo {author} {\bibfnamefont {Q.}~\bibnamefont {Wang}}, \bibinfo {author} {\bibfnamefont {L.~Y.}\ \bibnamefont {Liao}}, \bibinfo {author} {\bibfnamefont {W.~X.}\ \bibnamefont {Zhu}}, \bibinfo {author} {\bibfnamefont {X.~Z.}\ \bibnamefont {Chen}}, \bibinfo {author} {\bibfnamefont {F.}~\bibnamefont {Pan}}, \bibinfo {author} {\bibfnamefont {X.~L.}\ \bibnamefont {Fan}},\ and\ \bibinfo {author} {\bibfnamefont {C.}~\bibnamefont {Song}},\ }\bibfield  {title} {\bibinfo {title} {Observation of spin splitting torque in a collinear antiferromagnet {RuO}$_{2}$},\ }\href {https://doi.org/10.1103/PhysRevLett.128.197202} {\bibfield  {journal} {\bibinfo  {journal} {Phys. Rev. Lett.}\ }\textbf
  {\bibinfo {volume} {128}},\ \bibinfo {pages} {197202} (\bibinfo {year} {2022})}\BibitemShut {NoStop}%
\bibitem [{\citenamefont {Karube}\ \emph {et~al.}(2022)\citenamefont {Karube}, \citenamefont {Tanaka}, \citenamefont {Sugawara}, \citenamefont {Kadoguchi}, \citenamefont {Kohda},\ and\ \citenamefont {Nitta}}]{Karube2021}%
  \BibitemOpen
  \bibfield  {author} {\bibinfo {author} {\bibfnamefont {S.}~\bibnamefont {Karube}}, \bibinfo {author} {\bibfnamefont {T.}~\bibnamefont {Tanaka}}, \bibinfo {author} {\bibfnamefont {D.}~\bibnamefont {Sugawara}}, \bibinfo {author} {\bibfnamefont {N.}~\bibnamefont {Kadoguchi}}, \bibinfo {author} {\bibfnamefont {M.}~\bibnamefont {Kohda}},\ and\ \bibinfo {author} {\bibfnamefont {J.}~\bibnamefont {Nitta}},\ }\bibfield  {title} {\bibinfo {title} {Observation of spin-splitter torque in collinear antiferromagnetic {RuO}$_{2}$},\ }\href {https://doi.org/10.1103/PhysRevLett.129.137201} {\bibfield  {journal} {\bibinfo  {journal} {Phys. Rev. Lett.}\ }\textbf {\bibinfo {volume} {129}},\ \bibinfo {pages} {137201} (\bibinfo {year} {2022})}\BibitemShut {NoStop}%
\bibitem [{\citenamefont {Reichlová}\ \emph {et~al.}(2020)\citenamefont {Reichlová}, \citenamefont {Seeger}, \citenamefont {González-Hernández}, \citenamefont {Kounta}, \citenamefont {Schlitz}, \citenamefont {Kriegner}, \citenamefont {Ritzinger}, \citenamefont {Lammel}, \citenamefont {Leiviskä}, \citenamefont {Petříček}, \citenamefont {Doležal}, \citenamefont {Schmoranzerová}, \citenamefont {Bad'ura}, \citenamefont {Thomas}, \citenamefont {Baltz}, \citenamefont {Michez}, \citenamefont {Sinova}, \citenamefont {Goennenwein}, \citenamefont {Jungwirth},\ and\ \citenamefont {Šmejkal}}]{Helena2020}%
  \BibitemOpen
  \bibfield  {author} {\bibinfo {author} {\bibfnamefont {H.}~\bibnamefont {Reichlová}}, \bibinfo {author} {\bibfnamefont {R.~L.}\ \bibnamefont {Seeger}}, \bibinfo {author} {\bibfnamefont {R.}~\bibnamefont {González-Hernández}}, \bibinfo {author} {\bibfnamefont {I.}~\bibnamefont {Kounta}}, \bibinfo {author} {\bibfnamefont {R.}~\bibnamefont {Schlitz}}, \bibinfo {author} {\bibfnamefont {D.}~\bibnamefont {Kriegner}}, \bibinfo {author} {\bibfnamefont {P.}~\bibnamefont {Ritzinger}}, \bibinfo {author} {\bibfnamefont {M.}~\bibnamefont {Lammel}}, \bibinfo {author} {\bibfnamefont {M.}~\bibnamefont {Leiviskä}}, \bibinfo {author} {\bibfnamefont {V.}~\bibnamefont {Petříček}}, \bibinfo {author} {\bibfnamefont {P.}~\bibnamefont {Doležal}}, \bibinfo {author} {\bibfnamefont {E.}~\bibnamefont {Schmoranzerová}}, \bibinfo {author} {\bibfnamefont {A.}~\bibnamefont {Bad'ura}}, \bibinfo {author} {\bibfnamefont {A.}~\bibnamefont {Thomas}}, \bibinfo {author} {\bibfnamefont {V.}~\bibnamefont {Baltz}}, \bibinfo {author}
  {\bibfnamefont {L.}~\bibnamefont {Michez}}, \bibinfo {author} {\bibfnamefont {J.}~\bibnamefont {Sinova}}, \bibinfo {author} {\bibfnamefont {S.~T.~B.}\ \bibnamefont {Goennenwein}}, \bibinfo {author} {\bibfnamefont {T.}~\bibnamefont {Jungwirth}},\ and\ \bibinfo {author} {\bibfnamefont {L.}~\bibnamefont {Šmejkal}},\ }\bibfield  {title} {\bibinfo {title} {Macroscopic time reversal symmetry breaking by staggered spin-momentum interaction},\ }\bibfield  {journal} {\bibinfo  {journal} {arXiv 2012.15651}\ }\href {https://doi.org/10.48550/ARXIV.2012.15651} {10.48550/ARXIV.2012.15651} (\bibinfo {year} {2020})\BibitemShut {NoStop}%
\bibitem [{\citenamefont {Feng}\ \emph {et~al.}(2022)\citenamefont {Feng}, \citenamefont {Zhou}, \citenamefont {{\v S}mejkal}, \citenamefont {Wu}, \citenamefont {Zhu}, \citenamefont {Guo}, \citenamefont {Gonz{\'a}lez-Hern{\'a}ndez}, \citenamefont {Wang}, \citenamefont {Yan}, \citenamefont {Qin}, \citenamefont {Zhang}, \citenamefont {Wu}, \citenamefont {Chen}, \citenamefont {Meng}, \citenamefont {Liu}, \citenamefont {Xia}, \citenamefont {Sinova}, \citenamefont {Jungwirth},\ and\ \citenamefont {Liu}}]{Feng2020}%
  \BibitemOpen
  \bibfield  {author} {\bibinfo {author} {\bibfnamefont {Z.}~\bibnamefont {Feng}}, \bibinfo {author} {\bibfnamefont {X.}~\bibnamefont {Zhou}}, \bibinfo {author} {\bibfnamefont {L.}~\bibnamefont {{\v S}mejkal}}, \bibinfo {author} {\bibfnamefont {L.}~\bibnamefont {Wu}}, \bibinfo {author} {\bibfnamefont {Z.}~\bibnamefont {Zhu}}, \bibinfo {author} {\bibfnamefont {H.}~\bibnamefont {Guo}}, \bibinfo {author} {\bibfnamefont {R.}~\bibnamefont {Gonz{\'a}lez-Hern{\'a}ndez}}, \bibinfo {author} {\bibfnamefont {X.}~\bibnamefont {Wang}}, \bibinfo {author} {\bibfnamefont {H.}~\bibnamefont {Yan}}, \bibinfo {author} {\bibfnamefont {P.}~\bibnamefont {Qin}}, \bibinfo {author} {\bibfnamefont {X.}~\bibnamefont {Zhang}}, \bibinfo {author} {\bibfnamefont {H.}~\bibnamefont {Wu}}, \bibinfo {author} {\bibfnamefont {H.}~\bibnamefont {Chen}}, \bibinfo {author} {\bibfnamefont {Z.}~\bibnamefont {Meng}}, \bibinfo {author} {\bibfnamefont {L.}~\bibnamefont {Liu}}, \bibinfo {author} {\bibfnamefont {Z.}~\bibnamefont {Xia}}, \bibinfo {author}
  {\bibfnamefont {J.}~\bibnamefont {Sinova}}, \bibinfo {author} {\bibfnamefont {T.}~\bibnamefont {Jungwirth}},\ and\ \bibinfo {author} {\bibfnamefont {Z.}~\bibnamefont {Liu}},\ }\bibfield  {title} {\bibinfo {title} {An anomalous {H}all effect in altermagnetic ruthenium dioxide},\ }\href {https://doi.org/10.1038/s41928-022-00866-z} {\bibfield  {journal} {\bibinfo  {journal} {Nat. Electron.}\ }\textbf {\bibinfo {volume} {5}},\ \bibinfo {pages} {735} (\bibinfo {year} {2022})}\BibitemShut {NoStop}%
\bibitem [{\citenamefont {Gonzalez~Betancourt}\ \emph {et~al.}(2023)\citenamefont {Gonzalez~Betancourt}, \citenamefont {Zub\'a\ifmmode~\check{c}\else \v{c}\fi{}}, \citenamefont {Gonzalez-Hernandez}, \citenamefont {Geishendorf}, \citenamefont {\ifmmode \check{S}\else \v{S}\fi{}ob\'a\ifmmode~\check{n}\else \v{n}\fi{}}, \citenamefont {Springholz}, \citenamefont {Olejn\'{\i}k}, \citenamefont {\ifmmode~\check{S}\else \v{S}\fi{}mejkal}, \citenamefont {Sinova}, \citenamefont {Jungwirth}, \citenamefont {Goennenwein}, \citenamefont {Thomas}, \citenamefont {Reichlov\'a}, \citenamefont {\ifmmode~\check{Z}\else \v{Z}\fi{}elezn\'y},\ and\ \citenamefont {Kriegner}}]{Betancourt2021}%
  \BibitemOpen
  \bibfield  {author} {\bibinfo {author} {\bibfnamefont {R.~D.}\ \bibnamefont {Gonzalez~Betancourt}}, \bibinfo {author} {\bibfnamefont {J.}~\bibnamefont {Zub\'a\ifmmode~\check{c}\else \v{c}\fi{}}}, \bibinfo {author} {\bibfnamefont {R.}~\bibnamefont {Gonzalez-Hernandez}}, \bibinfo {author} {\bibfnamefont {K.}~\bibnamefont {Geishendorf}}, \bibinfo {author} {\bibfnamefont {Z.}~\bibnamefont {\ifmmode \check{S}\else \v{S}\fi{}ob\'a\ifmmode~\check{n}\else \v{n}\fi{}}}, \bibinfo {author} {\bibfnamefont {G.}~\bibnamefont {Springholz}}, \bibinfo {author} {\bibfnamefont {K.}~\bibnamefont {Olejn\'{\i}k}}, \bibinfo {author} {\bibfnamefont {L.}~\bibnamefont {\ifmmode~\check{S}\else \v{S}\fi{}mejkal}}, \bibinfo {author} {\bibfnamefont {J.}~\bibnamefont {Sinova}}, \bibinfo {author} {\bibfnamefont {T.}~\bibnamefont {Jungwirth}}, \bibinfo {author} {\bibfnamefont {S.~T.~B.}\ \bibnamefont {Goennenwein}}, \bibinfo {author} {\bibfnamefont {A.}~\bibnamefont {Thomas}}, \bibinfo {author} {\bibfnamefont {H.}~\bibnamefont
  {Reichlov\'a}}, \bibinfo {author} {\bibfnamefont {J.}~\bibnamefont {\ifmmode~\check{Z}\else \v{Z}\fi{}elezn\'y}},\ and\ \bibinfo {author} {\bibfnamefont {D.}~\bibnamefont {Kriegner}},\ }\bibfield  {title} {\bibinfo {title} {Spontaneous anomalous {H}all effect arising from an unconventional compensated magnetic phase in a semiconductor},\ }\href {https://doi.org/10.1103/PhysRevLett.130.036702} {\bibfield  {journal} {\bibinfo  {journal} {Phys. Rev. Lett.}\ }\textbf {\bibinfo {volume} {130}},\ \bibinfo {pages} {036702} (\bibinfo {year} {2023})}\BibitemShut {NoStop}%
\bibitem [{\citenamefont {{\v S}mejkal}\ \emph {et~al.}(2022)\citenamefont {{\v S}mejkal}, \citenamefont {MacDonald}, \citenamefont {Sinova}, \citenamefont {Nakatsuji},\ and\ \citenamefont {Jungwirth}}]{SmejkalAHE2022}%
  \BibitemOpen
  \bibfield  {author} {\bibinfo {author} {\bibfnamefont {L.}~\bibnamefont {{\v S}mejkal}}, \bibinfo {author} {\bibfnamefont {A.~H.}\ \bibnamefont {MacDonald}}, \bibinfo {author} {\bibfnamefont {J.}~\bibnamefont {Sinova}}, \bibinfo {author} {\bibfnamefont {S.}~\bibnamefont {Nakatsuji}},\ and\ \bibinfo {author} {\bibfnamefont {T.}~\bibnamefont {Jungwirth}},\ }\bibfield  {title} {\bibinfo {title} {Anomalous {H}all antiferromagnets},\ }\href {https://doi.org/10.1038/s41578-022-00430-3} {\bibfield  {journal} {\bibinfo  {journal} {Nat. Rev. Mater.}\ }\textbf {\bibinfo {volume} {7}},\ \bibinfo {pages} {482} (\bibinfo {year} {2022})}\BibitemShut {NoStop}%
\bibitem [{\citenamefont {\ifmmode~\check{S}\else \v{S}\fi{}mejkal}\ \emph {et~al.}(2023)\citenamefont {\ifmmode~\check{S}\else \v{S}\fi{}mejkal}, \citenamefont {Marmodoro}, \citenamefont {Ahn}, \citenamefont {Gonz\'alez-Hern\'andez}, \citenamefont {Turek}, \citenamefont {Mankovsky}, \citenamefont {Ebert}, \citenamefont {D'Souza}, \citenamefont {\ifmmode~\check{S}\else \v{S}\fi{}ipr}, \citenamefont {Sinova},\ and\ \citenamefont {Jungwirth}}]{Libor2023}%
  \BibitemOpen
  \bibfield  {author} {\bibinfo {author} {\bibfnamefont {L.}~\bibnamefont {\ifmmode~\check{S}\else \v{S}\fi{}mejkal}}, \bibinfo {author} {\bibfnamefont {A.}~\bibnamefont {Marmodoro}}, \bibinfo {author} {\bibfnamefont {K.-H.}\ \bibnamefont {Ahn}}, \bibinfo {author} {\bibfnamefont {R.}~\bibnamefont {Gonz\'alez-Hern\'andez}}, \bibinfo {author} {\bibfnamefont {I.}~\bibnamefont {Turek}}, \bibinfo {author} {\bibfnamefont {S.}~\bibnamefont {Mankovsky}}, \bibinfo {author} {\bibfnamefont {H.}~\bibnamefont {Ebert}}, \bibinfo {author} {\bibfnamefont {S.~W.}\ \bibnamefont {D'Souza}}, \bibinfo {author} {\bibfnamefont {O.~c.~v.}\ \bibnamefont {\ifmmode~\check{S}\else \v{S}\fi{}ipr}}, \bibinfo {author} {\bibfnamefont {J.}~\bibnamefont {Sinova}},\ and\ \bibinfo {author} {\bibfnamefont {T.~c.~v.}\ \bibnamefont {Jungwirth}},\ }\bibfield  {title} {\bibinfo {title} {Chiral magnons in altermagnetic {RuO}$_{2}$},\ }\href {https://doi.org/10.1103/PhysRevLett.131.256703} {\bibfield  {journal} {\bibinfo  {journal} {Phys. Rev. Lett.}\
  }\textbf {\bibinfo {volume} {131}},\ \bibinfo {pages} {256703} (\bibinfo {year} {2023})}\BibitemShut {NoStop}%
\bibitem [{\citenamefont {\ifmmode~\check{S}\else \v{S}\fi{}mejkal}\ \emph {et~al.}(2022{\natexlab{b}})\citenamefont {\ifmmode~\check{S}\else \v{S}\fi{}mejkal}, \citenamefont {Hellenes}, \citenamefont {Gonz\'alez-Hern\'andez}, \citenamefont {Sinova},\ and\ \citenamefont {Jungwirth}}]{Libor2022}%
  \BibitemOpen
  \bibfield  {author} {\bibinfo {author} {\bibfnamefont {L.}~\bibnamefont {\ifmmode~\check{S}\else \v{S}\fi{}mejkal}}, \bibinfo {author} {\bibfnamefont {A.~B.}\ \bibnamefont {Hellenes}}, \bibinfo {author} {\bibfnamefont {R.}~\bibnamefont {Gonz\'alez-Hern\'andez}}, \bibinfo {author} {\bibfnamefont {J.}~\bibnamefont {Sinova}},\ and\ \bibinfo {author} {\bibfnamefont {T.}~\bibnamefont {Jungwirth}},\ }\bibfield  {title} {\bibinfo {title} {Giant and tunneling magnetoresistance in unconventional collinear antiferromagnets with nonrelativistic spin-momentum coupling},\ }\href {https://doi.org/10.1103/PhysRevX.12.011028} {\bibfield  {journal} {\bibinfo  {journal} {Phys. Rev. X}\ }\textbf {\bibinfo {volume} {12}},\ \bibinfo {pages} {011028} (\bibinfo {year} {2022}{\natexlab{b}})}\BibitemShut {NoStop}%
\bibitem [{\citenamefont {Mazin}(2022)}]{Mazin2022}%
  \BibitemOpen
  \bibfield  {author} {\bibinfo {author} {\bibfnamefont {I.~I.}\ \bibnamefont {Mazin}},\ }\bibfield  {title} {\bibinfo {title} {Notes on altermagnetism and superconductivity},\ }\bibfield  {journal} {\bibinfo  {journal} {arXiv 2203.05000}\ }\href {https://doi.org/10.48550/ARXIV.2203.05000} {10.48550/ARXIV.2203.05000} (\bibinfo {year} {2022})\BibitemShut {NoStop}%
\bibitem [{\citenamefont {Zhu}\ \emph {et~al.}(2023)\citenamefont {Zhu}, \citenamefont {Zhuang}, \citenamefont {Wu},\ and\ \citenamefont {Yan}}]{Zhu2023}%
  \BibitemOpen
  \bibfield  {author} {\bibinfo {author} {\bibfnamefont {D.}~\bibnamefont {Zhu}}, \bibinfo {author} {\bibfnamefont {Z.-Y.}\ \bibnamefont {Zhuang}}, \bibinfo {author} {\bibfnamefont {Z.}~\bibnamefont {Wu}},\ and\ \bibinfo {author} {\bibfnamefont {Z.}~\bibnamefont {Yan}},\ }\bibfield  {title} {\bibinfo {title} {Topological superconductivity in two-dimensional altermagnetic metals},\ }\href {https://doi.org/10.1103/PhysRevB.108.184505} {\bibfield  {journal} {\bibinfo  {journal} {Phys. Rev. B}\ }\textbf {\bibinfo {volume} {108}},\ \bibinfo {pages} {184505} (\bibinfo {year} {2023})}\BibitemShut {NoStop}%
\bibitem [{\citenamefont {Zhang}\ \emph {et~al.}(2024)\citenamefont {Zhang}, \citenamefont {Hu},\ and\ \citenamefont {Neupert}}]{Zhang2024}%
  \BibitemOpen
  \bibfield  {author} {\bibinfo {author} {\bibfnamefont {S.-B.}\ \bibnamefont {Zhang}}, \bibinfo {author} {\bibfnamefont {L.-H.}\ \bibnamefont {Hu}},\ and\ \bibinfo {author} {\bibfnamefont {T.}~\bibnamefont {Neupert}},\ }\bibfield  {title} {\bibinfo {title} {Finite-momentum cooper pairing in proximitized altermagnets},\ }\href {https://doi.org/10.1038/s41467-024-45951-3} {\bibfield  {journal} {\bibinfo  {journal} {Nature Communications}\ }\textbf {\bibinfo {volume} {15}},\ \bibinfo {pages} {1801} (\bibinfo {year} {2024})}\BibitemShut {NoStop}%
\bibitem [{\citenamefont {Bhowal}\ and\ \citenamefont {Spaldin}(2024)}]{BhowalSpaldin2024}%
  \BibitemOpen
  \bibfield  {author} {\bibinfo {author} {\bibfnamefont {S.}~\bibnamefont {Bhowal}}\ and\ \bibinfo {author} {\bibfnamefont {N.~A.}\ \bibnamefont {Spaldin}},\ }\bibfield  {title} {\bibinfo {title} {Ferroically ordered magnetic octupoles in $d$-wave altermagnets},\ }\href {https://doi.org/10.1103/PhysRevX.14.011019} {\bibfield  {journal} {\bibinfo  {journal} {Phys. Rev. X}\ }\textbf {\bibinfo {volume} {14}},\ \bibinfo {pages} {011019} (\bibinfo {year} {2024})}\BibitemShut {NoStop}%
\bibitem [{\citenamefont {Lovesey}\ \emph {et~al.}(2023)\citenamefont {Lovesey}, \citenamefont {Khalyavin},\ and\ \citenamefont {van~der Laan}}]{vanderLaan2023}%
  \BibitemOpen
  \bibfield  {author} {\bibinfo {author} {\bibfnamefont {S.~W.}\ \bibnamefont {Lovesey}}, \bibinfo {author} {\bibfnamefont {D.~D.}\ \bibnamefont {Khalyavin}},\ and\ \bibinfo {author} {\bibfnamefont {G.}~\bibnamefont {van~der Laan}},\ }\bibfield  {title} {\bibinfo {title} {Templates for magnetic symmetry and altermagnetism in hexagonal {MnTe}},\ }\href {https://doi.org/10.1103/PhysRevB.108.174437} {\bibfield  {journal} {\bibinfo  {journal} {Phys. Rev. B}\ }\textbf {\bibinfo {volume} {108}},\ \bibinfo {pages} {174437} (\bibinfo {year} {2023})}\BibitemShut {NoStop}%
\bibitem [{\citenamefont {Lee}\ \emph {et~al.}(2024)\citenamefont {Lee}, \citenamefont {Lee}, \citenamefont {Jung}, \citenamefont {Jung}, \citenamefont {Kim}, \citenamefont {Lee}, \citenamefont {Seok}, \citenamefont {Kim}, \citenamefont {Park}, \citenamefont {\ifmmode~\check{S}\else \v{S}\fi{}mejkal}, \citenamefont {Kang},\ and\ \citenamefont {Kim}}]{Lee2024}%
  \BibitemOpen
  \bibfield  {author} {\bibinfo {author} {\bibfnamefont {S.}~\bibnamefont {Lee}}, \bibinfo {author} {\bibfnamefont {S.}~\bibnamefont {Lee}}, \bibinfo {author} {\bibfnamefont {S.}~\bibnamefont {Jung}}, \bibinfo {author} {\bibfnamefont {J.}~\bibnamefont {Jung}}, \bibinfo {author} {\bibfnamefont {D.}~\bibnamefont {Kim}}, \bibinfo {author} {\bibfnamefont {Y.}~\bibnamefont {Lee}}, \bibinfo {author} {\bibfnamefont {B.}~\bibnamefont {Seok}}, \bibinfo {author} {\bibfnamefont {J.}~\bibnamefont {Kim}}, \bibinfo {author} {\bibfnamefont {B.~G.}\ \bibnamefont {Park}}, \bibinfo {author} {\bibfnamefont {L.}~\bibnamefont {\ifmmode~\check{S}\else \v{S}\fi{}mejkal}}, \bibinfo {author} {\bibfnamefont {C.-J.}\ \bibnamefont {Kang}},\ and\ \bibinfo {author} {\bibfnamefont {C.}~\bibnamefont {Kim}},\ }\bibfield  {title} {\bibinfo {title} {Broken {K}ramers degeneracy in altermagnetic {MnTe}},\ }\href {https://doi.org/10.1103/PhysRevLett.132.036702} {\bibfield  {journal} {\bibinfo  {journal} {Phys. Rev. Lett.}\ }\textbf {\bibinfo
  {volume} {132}},\ \bibinfo {pages} {036702} (\bibinfo {year} {2024})}\BibitemShut {NoStop}%
\bibitem [{\citenamefont {Krempask{\'y}}\ \emph {et~al.}(2024)\citenamefont {Krempask{\'y}}, \citenamefont {{\v S}mejkal}, \citenamefont {D'Souza}, \citenamefont {Hajlaoui}, \citenamefont {Springholz}, \citenamefont {Uhl{\'\i}{\v r}ov{\'a}}, \citenamefont {Alarab}, \citenamefont {Constantinou}, \citenamefont {Strocov}, \citenamefont {Usanov}, \citenamefont {Pudelko}, \citenamefont {Gonz{\'a}lez-Hern{\'a}ndez}, \citenamefont {Birk~Hellenes}, \citenamefont {Jansa}, \citenamefont {Reichlov{\'a}}, \citenamefont {{\v S}ob{\'a}{\v n}}, \citenamefont {Gonzalez~Betancourt}, \citenamefont {Wadley}, \citenamefont {Sinova}, \citenamefont {Kriegner}, \citenamefont {Min{\'a}r}, \citenamefont {Dil},\ and\ \citenamefont {Jungwirth}}]{Krempask2024}%
  \BibitemOpen
  \bibfield  {author} {\bibinfo {author} {\bibfnamefont {J.}~\bibnamefont {Krempask{\'y}}}, \bibinfo {author} {\bibfnamefont {L.}~\bibnamefont {{\v S}mejkal}}, \bibinfo {author} {\bibfnamefont {S.~W.}\ \bibnamefont {D'Souza}}, \bibinfo {author} {\bibfnamefont {M.}~\bibnamefont {Hajlaoui}}, \bibinfo {author} {\bibfnamefont {G.}~\bibnamefont {Springholz}}, \bibinfo {author} {\bibfnamefont {K.}~\bibnamefont {Uhl{\'\i}{\v r}ov{\'a}}}, \bibinfo {author} {\bibfnamefont {F.}~\bibnamefont {Alarab}}, \bibinfo {author} {\bibfnamefont {P.~C.}\ \bibnamefont {Constantinou}}, \bibinfo {author} {\bibfnamefont {V.}~\bibnamefont {Strocov}}, \bibinfo {author} {\bibfnamefont {D.}~\bibnamefont {Usanov}}, \bibinfo {author} {\bibfnamefont {W.~R.}\ \bibnamefont {Pudelko}}, \bibinfo {author} {\bibfnamefont {R.}~\bibnamefont {Gonz{\'a}lez-Hern{\'a}ndez}}, \bibinfo {author} {\bibfnamefont {A.}~\bibnamefont {Birk~Hellenes}}, \bibinfo {author} {\bibfnamefont {Z.}~\bibnamefont {Jansa}}, \bibinfo {author} {\bibfnamefont {H.}~\bibnamefont
  {Reichlov{\'a}}}, \bibinfo {author} {\bibfnamefont {Z.}~\bibnamefont {{\v S}ob{\'a}{\v n}}}, \bibinfo {author} {\bibfnamefont {R.~D.}\ \bibnamefont {Gonzalez~Betancourt}}, \bibinfo {author} {\bibfnamefont {P.}~\bibnamefont {Wadley}}, \bibinfo {author} {\bibfnamefont {J.}~\bibnamefont {Sinova}}, \bibinfo {author} {\bibfnamefont {D.}~\bibnamefont {Kriegner}}, \bibinfo {author} {\bibfnamefont {J.}~\bibnamefont {Min{\'a}r}}, \bibinfo {author} {\bibfnamefont {J.~H.}\ \bibnamefont {Dil}},\ and\ \bibinfo {author} {\bibfnamefont {T.}~\bibnamefont {Jungwirth}},\ }\bibfield  {title} {\bibinfo {title} {Altermagnetic lifting of {K}ramers spin degeneracy},\ }\href {https://doi.org/10.1038/s41586-023-06907-7} {\bibfield  {journal} {\bibinfo  {journal} {Nature}\ }\textbf {\bibinfo {volume} {626}},\ \bibinfo {pages} {517} (\bibinfo {year} {2024})}\BibitemShut {NoStop}%
\bibitem [{\citenamefont {Reimers}\ \emph {et~al.}(2024)\citenamefont {Reimers}, \citenamefont {Odenbreit}, \citenamefont {Šmejkal}, \citenamefont {Strocov}, \citenamefont {Constantinou}, \citenamefont {Hellenes}, \citenamefont {Jaeschke~Ubiergo}, \citenamefont {Campos}, \citenamefont {Bharadwaj}, \citenamefont {Chakraborty}, \citenamefont {Denneulin}, \citenamefont {Shi}, \citenamefont {Dunin-Borkowski}, \citenamefont {Das}, \citenamefont {Kläui}, \citenamefont {Sinova},\ and\ \citenamefont {Jourdan}}]{reimers_direct_2024}%
  \BibitemOpen
  \bibfield  {author} {\bibinfo {author} {\bibfnamefont {S.}~\bibnamefont {Reimers}}, \bibinfo {author} {\bibfnamefont {L.}~\bibnamefont {Odenbreit}}, \bibinfo {author} {\bibfnamefont {L.}~\bibnamefont {Šmejkal}}, \bibinfo {author} {\bibfnamefont {V.~N.}\ \bibnamefont {Strocov}}, \bibinfo {author} {\bibfnamefont {P.}~\bibnamefont {Constantinou}}, \bibinfo {author} {\bibfnamefont {A.~B.}\ \bibnamefont {Hellenes}}, \bibinfo {author} {\bibfnamefont {R.}~\bibnamefont {Jaeschke~Ubiergo}}, \bibinfo {author} {\bibfnamefont {W.~H.}\ \bibnamefont {Campos}}, \bibinfo {author} {\bibfnamefont {V.~K.}\ \bibnamefont {Bharadwaj}}, \bibinfo {author} {\bibfnamefont {A.}~\bibnamefont {Chakraborty}}, \bibinfo {author} {\bibfnamefont {T.}~\bibnamefont {Denneulin}}, \bibinfo {author} {\bibfnamefont {W.}~\bibnamefont {Shi}}, \bibinfo {author} {\bibfnamefont {R.~E.}\ \bibnamefont {Dunin-Borkowski}}, \bibinfo {author} {\bibfnamefont {S.}~\bibnamefont {Das}}, \bibinfo {author} {\bibfnamefont {M.}~\bibnamefont {Kläui}}, \bibinfo
  {author} {\bibfnamefont {J.}~\bibnamefont {Sinova}},\ and\ \bibinfo {author} {\bibfnamefont {M.}~\bibnamefont {Jourdan}},\ }\bibfield  {title} {\bibinfo {title} {Direct observation of altermagnetic band splitting in {CrSb} thin films},\ }\href {https://doi.org/10.1038/s41467-024-46476-5} {\bibfield  {journal} {\bibinfo  {journal} {Nat Commun}\ }\textbf {\bibinfo {volume} {15}},\ \bibinfo {pages} {2116} (\bibinfo {year} {2024})}\BibitemShut {NoStop}%
\bibitem [{\citenamefont {Morin}(1951)}]{Morin1951}%
  \BibitemOpen
  \bibfield  {author} {\bibinfo {author} {\bibfnamefont {F.~J.}\ \bibnamefont {Morin}},\ }\bibfield  {title} {\bibinfo {title} {Electrical properties of $\alpha$-{Fe}$_2${O}$_3$ and $\alpha$-{Fe}$_2${O}$_3$ containing titanium},\ }\href {https://doi.org/10.1103/PhysRev.83.1005} {\bibfield  {journal} {\bibinfo  {journal} {Phys. Rev.}\ }\textbf {\bibinfo {volume} {83}},\ \bibinfo {pages} {1005} (\bibinfo {year} {1951})}\BibitemShut {NoStop}%
\bibitem [{\citenamefont {Dzyaloshinsky}(1958)}]{Dzyaloshinsky1958}%
  \BibitemOpen
  \bibfield  {author} {\bibinfo {author} {\bibfnamefont {I.}~\bibnamefont {Dzyaloshinsky}},\ }\bibfield  {title} {\bibinfo {title} {A thermodynamic theory of “weak” ferromagnetism of antiferromagnetics},\ }\href {https://doi.org/https://doi.org/10.1016/0022-3697(58)90076-3} {\bibfield  {journal} {\bibinfo  {journal} {J. Phys. Chem. Solids}\ }\textbf {\bibinfo {volume} {4}},\ \bibinfo {pages} {241} (\bibinfo {year} {1958})}\BibitemShut {NoStop}%
\bibitem [{\citenamefont {Kresse}\ and\ \citenamefont {Furthmüller}(1996{\natexlab{a}})}]{kresse_efficient_1996}%
  \BibitemOpen
  \bibfield  {author} {\bibinfo {author} {\bibfnamefont {G.}~\bibnamefont {Kresse}}\ and\ \bibinfo {author} {\bibfnamefont {J.}~\bibnamefont {Furthmüller}},\ }\bibfield  {title} {\bibinfo {title} {Efficient iterative schemes for ab initio total-energy calculations using a plane-wave basis set},\ }\bibfield  {journal} {\bibinfo  {journal} {Phys. Rev. B}\ }\textbf {\bibinfo {volume} {54}},\ \href {https://doi.org/10.1103/PhysRevB.54.11169} {10.1103/PhysRevB.54.11169} (\bibinfo {year} {1996}{\natexlab{a}})\BibitemShut {NoStop}%
\bibitem [{\citenamefont {Kresse}\ and\ \citenamefont {Furthmüller}(1996{\natexlab{b}})}]{kresse_efficiency_1996}%
  \BibitemOpen
  \bibfield  {author} {\bibinfo {author} {\bibfnamefont {G.}~\bibnamefont {Kresse}}\ and\ \bibinfo {author} {\bibfnamefont {J.}~\bibnamefont {Furthmüller}},\ }\bibfield  {title} {\bibinfo {title} {Efficiency of ab-initio total energy calculations for metals and semiconductors using a plane-wave basis set},\ }\bibfield  {journal} {\bibinfo  {journal} {Comp. Mater. Sci.}\ }\textbf {\bibinfo {volume} {6}},\ \href {https://doi.org/10.1016/0927-0256(96)00008-0} {10.1016/0927-0256(96)00008-0} (\bibinfo {year} {1996}{\natexlab{b}})\BibitemShut {NoStop}%
\bibitem [{\citenamefont {Perdew}\ and\ \citenamefont {Zunger}(1981)}]{LDA_pz}%
  \BibitemOpen
  \bibfield  {author} {\bibinfo {author} {\bibfnamefont {J.~P.}\ \bibnamefont {Perdew}}\ and\ \bibinfo {author} {\bibfnamefont {A.}~\bibnamefont {Zunger}},\ }\bibfield  {title} {\bibinfo {title} {{Self-interaction correction to density-functional approximations for many-electron systems}},\ }\href {https://doi.org/10.1103/PhysRevB.23.5048} {\bibfield  {journal} {\bibinfo  {journal} {Phys. Rev. B}\ }\textbf {\bibinfo {volume} {23}},\ \bibinfo {pages} {5048} (\bibinfo {year} {1981})}\BibitemShut {NoStop}%
\bibitem [{\citenamefont {Liechtenstein}\ \emph {et~al.}(1995)\citenamefont {Liechtenstein}, \citenamefont {Anisimov},\ and\ \citenamefont {Zaanen}}]{Liechtenstein_U}%
  \BibitemOpen
  \bibfield  {author} {\bibinfo {author} {\bibfnamefont {A.~I.}\ \bibnamefont {Liechtenstein}}, \bibinfo {author} {\bibfnamefont {V.~I.}\ \bibnamefont {Anisimov}},\ and\ \bibinfo {author} {\bibfnamefont {J.}~\bibnamefont {Zaanen}},\ }\bibfield  {title} {\bibinfo {title} {{Density-functional theory and strong interactions: Orbital ordering in Mott-Hubbard insulators}},\ }\href {https://doi.org/10.1103/PhysRevB.52.R5467} {\bibfield  {journal} {\bibinfo  {journal} {Phys. Rev. B}\ }\textbf {\bibinfo {volume} {52}},\ \bibinfo {pages} {R5467} (\bibinfo {year} {1995})}\BibitemShut {NoStop}%
\bibitem [{\citenamefont {Blöchl}(1994)}]{blochl_projector_1994}%
  \BibitemOpen
  \bibfield  {author} {\bibinfo {author} {\bibfnamefont {P.~E.}\ \bibnamefont {Blöchl}},\ }\bibfield  {title} {\bibinfo {title} {Projector augmented-wave method},\ }\href {https://doi.org/10.1103/PhysRevB.50.17953} {\bibfield  {journal} {\bibinfo  {journal} {Phys. Rev. B}\ }\textbf {\bibinfo {volume} {50}},\ \bibinfo {pages} {17953} (\bibinfo {year} {1994})}\BibitemShut {NoStop}%
\bibitem [{\citenamefont {Baron}\ \emph {et~al.}(2005)\citenamefont {Baron}, \citenamefont {Gutzmer}, \citenamefont {Rundlöf},\ and\ \citenamefont {Tellgren}}]{baron_neutron_2005}%
  \BibitemOpen
  \bibfield  {author} {\bibinfo {author} {\bibfnamefont {V.}~\bibnamefont {Baron}}, \bibinfo {author} {\bibfnamefont {J.}~\bibnamefont {Gutzmer}}, \bibinfo {author} {\bibfnamefont {H.}~\bibnamefont {Rundlöf}},\ and\ \bibinfo {author} {\bibfnamefont {R.}~\bibnamefont {Tellgren}},\ }\bibfield  {title} {\bibinfo {title} {Neutron powder diffraction study of {Mn}-bearing hematite, $\alpha$-{Fe}$_{2-x}$-{Mn}$_x${O}$_3$, in the range 0 $\leq$ x $\leq$ 0.176},\ }\href {https://doi.org/10.1016/j.solidstatesciences.2004.11.021} {\bibfield  {journal} {\bibinfo  {journal} {Solid State Sci.}\ }\textbf {\bibinfo {volume} {7}},\ \bibinfo {pages} {753} (\bibinfo {year} {2005})}\BibitemShut {NoStop}%
\bibitem [{\citenamefont {Hill}\ \emph {et~al.}(2008{\natexlab{a}})\citenamefont {Hill}, \citenamefont {Jiao}, \citenamefont {Bruce}, \citenamefont {Harrison}, \citenamefont {Kockelmann},\ and\ \citenamefont {Ritter}}]{hillNeutronDiffractionStudy2008}%
  \BibitemOpen
  \bibfield  {author} {\bibinfo {author} {\bibfnamefont {A.~H.}\ \bibnamefont {Hill}}, \bibinfo {author} {\bibfnamefont {F.}~\bibnamefont {Jiao}}, \bibinfo {author} {\bibfnamefont {P.~G.}\ \bibnamefont {Bruce}}, \bibinfo {author} {\bibfnamefont {A.}~\bibnamefont {Harrison}}, \bibinfo {author} {\bibfnamefont {W.}~\bibnamefont {Kockelmann}},\ and\ \bibinfo {author} {\bibfnamefont {C.}~\bibnamefont {Ritter}},\ }\bibfield  {title} {\bibinfo {title} {Neutron diffraction study of mesoporous and bulk hematite, $\alpha$-{Fe}$_2${O}$_3$},\ }\href {https://doi.org/10.1021/cm800009s} {\bibfield  {journal} {\bibinfo  {journal} {Chem. Mater.}\ }\textbf {\bibinfo {volume} {20}},\ \bibinfo {pages} {4891} (\bibinfo {year} {2008}{\natexlab{a}})}\BibitemShut {NoStop}%
\bibitem [{\citenamefont {Mochizuki}(1977)}]{mochizuki_electrical_1977}%
  \BibitemOpen
  \bibfield  {author} {\bibinfo {author} {\bibfnamefont {S.}~\bibnamefont {Mochizuki}},\ }\bibfield  {title} {\bibinfo {title} {Electrical conductivity of $\alpha$-{Fe}$_2${O}$_3$},\ }\href {https://doi.org/10.1002/pssa.2210410232} {\bibfield  {journal} {\bibinfo  {journal} {physica status solidi (a)}\ }\textbf {\bibinfo {volume} {41}},\ \bibinfo {pages} {591} (\bibinfo {year} {1977})}\BibitemShut {NoStop}%
\bibitem [{\citenamefont {Gilbert}\ \emph {et~al.}(2009)\citenamefont {Gilbert}, \citenamefont {Frandsen}, \citenamefont {Maxey},\ and\ \citenamefont {Sherman}}]{gilbert_band-gap_2009}%
  \BibitemOpen
  \bibfield  {author} {\bibinfo {author} {\bibfnamefont {B.}~\bibnamefont {Gilbert}}, \bibinfo {author} {\bibfnamefont {C.}~\bibnamefont {Frandsen}}, \bibinfo {author} {\bibfnamefont {E.~R.}\ \bibnamefont {Maxey}},\ and\ \bibinfo {author} {\bibfnamefont {D.~M.}\ \bibnamefont {Sherman}},\ }\bibfield  {title} {\bibinfo {title} {Band-gap measurements of bulk and nanoscale hematite by soft x-ray spectroscopy},\ }\href {https://doi.org/10.1103/PhysRevB.79.035108} {\bibfield  {journal} {\bibinfo  {journal} {Phys. Rev. B}\ }\textbf {\bibinfo {volume} {79}},\ \bibinfo {pages} {035108} (\bibinfo {year} {2009})}\BibitemShut {NoStop}%
\bibitem [{\citenamefont {Verbeek}\ \emph {et~al.}(2023)\citenamefont {Verbeek}, \citenamefont {Urru},\ and\ \citenamefont {Spaldin}}]{Verbeek_Hidden_2023}%
  \BibitemOpen
  \bibfield  {author} {\bibinfo {author} {\bibfnamefont {X.~H.}\ \bibnamefont {Verbeek}}, \bibinfo {author} {\bibfnamefont {A.}~\bibnamefont {Urru}},\ and\ \bibinfo {author} {\bibfnamefont {N.~A.}\ \bibnamefont {Spaldin}},\ }\bibfield  {title} {\bibinfo {title} {Hidden orders and (anti-)magnetoelectric effects in {Cr}$_2${O}$_{3}$ and $\alpha$-{Fe}$_2${O}$_{3}$},\ }\href {https://doi.org/10.1103/PhysRevResearch.5.L042018} {\bibfield  {journal} {\bibinfo  {journal} {Phys. Rev. Res.}\ }\textbf {\bibinfo {volume} {5}},\ \bibinfo {pages} {L042018} (\bibinfo {year} {2023})}\BibitemShut {NoStop}%
\bibitem [{\citenamefont {Hill}\ \emph {et~al.}(2008{\natexlab{b}})\citenamefont {Hill}, \citenamefont {Jiao}, \citenamefont {Bruce}, \citenamefont {Harrison}, \citenamefont {Kockelmann},\ and\ \citenamefont {Ritter}}]{hill_neutron_2008}%
  \BibitemOpen
  \bibfield  {author} {\bibinfo {author} {\bibfnamefont {A.~H.}\ \bibnamefont {Hill}}, \bibinfo {author} {\bibfnamefont {F.}~\bibnamefont {Jiao}}, \bibinfo {author} {\bibfnamefont {P.~G.}\ \bibnamefont {Bruce}}, \bibinfo {author} {\bibfnamefont {A.}~\bibnamefont {Harrison}}, \bibinfo {author} {\bibfnamefont {W.}~\bibnamefont {Kockelmann}},\ and\ \bibinfo {author} {\bibfnamefont {C.}~\bibnamefont {Ritter}},\ }\bibfield  {title} {\bibinfo {title} {Neutron diffraction study of mesoporous and bulk hematite, $\alpha$-{Fe}$_2${O}$_3$},\ }\href {https://doi.org/10.1021/cm800009s} {\bibfield  {journal} {\bibinfo  {journal} {Chem. Mater.}\ }\textbf {\bibinfo {volume} {20}},\ \bibinfo {pages} {4891} (\bibinfo {year} {2008}{\natexlab{b}})}\BibitemShut {NoStop}%
\bibitem [{\citenamefont {Ma}\ and\ \citenamefont {Dudarev}(2015)}]{ma_constrained_2015}%
  \BibitemOpen
  \bibfield  {author} {\bibinfo {author} {\bibfnamefont {P.-W.}\ \bibnamefont {Ma}}\ and\ \bibinfo {author} {\bibfnamefont {S.~L.}\ \bibnamefont {Dudarev}},\ }\bibfield  {title} {\bibinfo {title} {Constrained density functional for noncollinear magnetism},\ }\href {https://doi.org/10.1103/PhysRevB.91.054420} {\bibfield  {journal} {\bibinfo  {journal} {Phys. Rev. B}\ }\textbf {\bibinfo {volume} {91}},\ \bibinfo {pages} {054420} (\bibinfo {year} {2015})}\BibitemShut {NoStop}%
\bibitem [{\citenamefont {Cricchio}(2010)}]{cricchio_multipoles_2010}%
  \BibitemOpen
  \bibfield  {author} {\bibinfo {author} {\bibfnamefont {F.}~\bibnamefont {Cricchio}},\ }\emph {\bibinfo {title} {Multipoles in {Correlated} {Electron} {Materials}}},\ \href {http://urn.kb.se/resolve?urn=urn:nbn:se:uu:diva-132068} {Ph.D. thesis},\ \bibinfo  {school} {Uppsala University} (\bibinfo {year} {2010})\BibitemShut {NoStop}%
\bibitem [{\citenamefont {Grånäs}(2012)}]{granas_theoretical_2012}%
  \BibitemOpen
  \bibfield  {author} {\bibinfo {author} {\bibfnamefont {O.}~\bibnamefont {Grånäs}},\ }\emph {\bibinfo {title} {Theoretical {Studies} of {Magnetism} and {Electron} {Correlation} in {Solids}}},\ \href {http://urn.kb.se/resolve?urn=urn:nbn:se:uu:diva-172334} {Ph.D. thesis},\ \bibinfo  {school} {Uppsala University} (\bibinfo {year} {2012})\BibitemShut {NoStop}%
\bibitem [{\citenamefont {Spaldin}\ \emph {et~al.}(2013)\citenamefont {Spaldin}, \citenamefont {Fechner}, \citenamefont {Bousquet}, \citenamefont {Balatsky},\ and\ \citenamefont {Nordström}}]{spaldin_monopole-based_2013}%
  \BibitemOpen
  \bibfield  {author} {\bibinfo {author} {\bibfnamefont {N.~A.}\ \bibnamefont {Spaldin}}, \bibinfo {author} {\bibfnamefont {M.}~\bibnamefont {Fechner}}, \bibinfo {author} {\bibfnamefont {E.}~\bibnamefont {Bousquet}}, \bibinfo {author} {\bibfnamefont {A.}~\bibnamefont {Balatsky}},\ and\ \bibinfo {author} {\bibfnamefont {L.}~\bibnamefont {Nordström}},\ }\bibfield  {title} {\bibinfo {title} {Monopole-based formalism for the diagonal magnetoelectric response},\ }\href {https://doi.org/10.1103/PhysRevB.88.094429} {\bibfield  {journal} {\bibinfo  {journal} {Phys. Rev. B}\ }\textbf {\bibinfo {volume} {88}},\ \bibinfo {pages} {094429} (\bibinfo {year} {2013})}\BibitemShut {NoStop}%
\bibitem [{\citenamefont {Merkel}(2023)}]{multipyles}%
  \BibitemOpen
  \bibfield  {author} {\bibinfo {author} {\bibfnamefont {M.~E.}\ \bibnamefont {Merkel}},\ }\href {https://doi.org/10.5281/zenodo.8199391} {\bibinfo {title} {multipyles v1.1.0}} (\bibinfo {year} {2023})\BibitemShut {NoStop}%
\bibitem [{\citenamefont {Schaufelberger}\ \emph {et~al.}(2023)\citenamefont {Schaufelberger}, \citenamefont {Merkel}, \citenamefont {Tehrani}, \citenamefont {Spaldin},\ and\ \citenamefont {Ederer}}]{schaufelberger_exploring_2023}%
  \BibitemOpen
  \bibfield  {author} {\bibinfo {author} {\bibfnamefont {L.}~\bibnamefont {Schaufelberger}}, \bibinfo {author} {\bibfnamefont {M.~E.}\ \bibnamefont {Merkel}}, \bibinfo {author} {\bibfnamefont {A.~M.}\ \bibnamefont {Tehrani}}, \bibinfo {author} {\bibfnamefont {N.~A.}\ \bibnamefont {Spaldin}},\ and\ \bibinfo {author} {\bibfnamefont {C.}~\bibnamefont {Ederer}},\ }\bibfield  {title} {\bibinfo {title} {Exploring energy landscapes of charge multipoles using constrained density functional theory},\ }\href {https://doi.org/10.1103/PhysRevResearch.5.033172} {\bibfield  {journal} {\bibinfo  {journal} {Phys. Rev. Res.}\ }\textbf {\bibinfo {volume} {5}},\ \bibinfo {pages} {033172} (\bibinfo {year} {2023})}\BibitemShut {NoStop}%
\bibitem [{\citenamefont {Cricchio}\ \emph {et~al.}(2009)\citenamefont {Cricchio}, \citenamefont {Bultmark}, \citenamefont {Grånäs},\ and\ \citenamefont {Nordström}}]{cricchio_itinerant_2009}%
  \BibitemOpen
  \bibfield  {author} {\bibinfo {author} {\bibfnamefont {F.}~\bibnamefont {Cricchio}}, \bibinfo {author} {\bibfnamefont {F.}~\bibnamefont {Bultmark}}, \bibinfo {author} {\bibfnamefont {O.}~\bibnamefont {Grånäs}},\ and\ \bibinfo {author} {\bibfnamefont {L.}~\bibnamefont {Nordström}},\ }\bibfield  {title} {\bibinfo {title} {Itinerant magnetic multipole moments of rank five as the hidden order in {URu}$_2${Si}$_2$},\ }\href {https://doi.org/10.1103/PhysRevLett.103.107202} {\bibfield  {journal} {\bibinfo  {journal} {Phys. Rev. Lett.}\ }\textbf {\bibinfo {volume} {103}},\ \bibinfo {pages} {107202} (\bibinfo {year} {2009})}\BibitemShut {NoStop}%
\bibitem [{\citenamefont {Bultmark}\ \emph {et~al.}(2009)\citenamefont {Bultmark}, \citenamefont {Cricchio}, \citenamefont {Gr{\aa}n{\"{a}}s},\ and\ \citenamefont {Nordstr{\"{o}}m}}]{multipole_decomposition}%
  \BibitemOpen
  \bibfield  {author} {\bibinfo {author} {\bibfnamefont {F.}~\bibnamefont {Bultmark}}, \bibinfo {author} {\bibfnamefont {F.}~\bibnamefont {Cricchio}}, \bibinfo {author} {\bibfnamefont {O.}~\bibnamefont {Gr{\aa}n{\"{a}}s}},\ and\ \bibinfo {author} {\bibfnamefont {L.}~\bibnamefont {Nordstr{\"{o}}m}},\ }\bibfield  {title} {\bibinfo {title} {{Multipole decomposition of LDA+U energy and its application to actinide compounds}},\ }\href {https://doi.org/10.1103/PhysRevB.80.035121} {\bibfield  {journal} {\bibinfo  {journal} {Phys. Rev. B}\ }\textbf {\bibinfo {volume} {80}},\ \bibinfo {pages} {035121} (\bibinfo {year} {2009})}\BibitemShut {NoStop}%
\bibitem [{\citenamefont {Samuelsen}\ \emph {et~al.}(1970)\citenamefont {Samuelsen}, \citenamefont {Hutchings},\ and\ \citenamefont {Shirane}}]{samuelsenInelasticNeutronScattering1970a}%
  \BibitemOpen
  \bibfield  {author} {\bibinfo {author} {\bibfnamefont {E.~J.}\ \bibnamefont {Samuelsen}}, \bibinfo {author} {\bibfnamefont {M.~T.}\ \bibnamefont {Hutchings}},\ and\ \bibinfo {author} {\bibfnamefont {G.}~\bibnamefont {Shirane}},\ }\bibfield  {title} {\bibinfo {title} {Inelastic neutron scattering investigation of spin waves and magnetic interactions in {{Cr$_2$O$_3$}}},\ }\href {https://doi.org/10.1016/0031-8914(70)90158-8} {\bibfield  {journal} {\bibinfo  {journal} {Physica}\ }\textbf {\bibinfo {volume} {48}},\ \bibinfo {pages} {13} (\bibinfo {year} {1970})}\BibitemShut {NoStop}%
\bibitem [{\citenamefont {Brown}\ \emph {et~al.}(2002)\citenamefont {Brown}, \citenamefont {Forsyth}, \citenamefont {{Leli{\`e}vre-Berna}},\ and\ \citenamefont {Tasset}}]{brownDeterminationMagnetizationDistribution2002}%
  \BibitemOpen
  \bibfield  {author} {\bibinfo {author} {\bibfnamefont {P.~J.}\ \bibnamefont {Brown}}, \bibinfo {author} {\bibfnamefont {J.~B.}\ \bibnamefont {Forsyth}}, \bibinfo {author} {\bibfnamefont {E.}~\bibnamefont {{Leli{\`e}vre-Berna}}},\ and\ \bibinfo {author} {\bibfnamefont {F.}~\bibnamefont {Tasset}},\ }\bibfield  {title} {\bibinfo {title} {Determination of the magnetization distribution in {Cr}$_2${O}$_3$ spherical neutron polarimetry},\ }\bibfield  {journal} {\bibinfo  {journal} {J. Phys. : Condens. Matter}\ }\textbf {\bibinfo {volume} {14}},\ \href {https://doi.org/10.1088/0953-8984/14/8/323} {10.1088/0953-8984/14/8/323} (\bibinfo {year} {2002})\BibitemShut {NoStop}%
\bibitem [{\citenamefont {Flanders}\ and\ \citenamefont {Remeika}(1965)}]{flanders_magnetic_1965}%
  \BibitemOpen
  \bibfield  {author} {\bibinfo {author} {\bibfnamefont {P.~J.}\ \bibnamefont {Flanders}}\ and\ \bibinfo {author} {\bibfnamefont {J.~P.}\ \bibnamefont {Remeika}},\ }\bibfield  {title} {\bibinfo {title} {Magnetic properties of hematite single crystals},\ }\href {https://doi.org/10.1080/14786436508224935} {\bibfield  {journal} {\bibinfo  {journal} {Phil. Mag: J. Theor. Exp. Appl. Phys.}\ }\textbf {\bibinfo {volume} {11}},\ \bibinfo {pages} {1271} (\bibinfo {year} {1965})}\BibitemShut {NoStop}%
\bibitem [{\citenamefont {Bødker}\ \emph {et~al.}(2000)\citenamefont {Bødker}, \citenamefont {Hansen}, \citenamefont {Koch}, \citenamefont {Lefmann},\ and\ \citenamefont {Mørup}}]{bodker_magnetic_2000}%
  \BibitemOpen
  \bibfield  {author} {\bibinfo {author} {\bibfnamefont {F.}~\bibnamefont {Bødker}}, \bibinfo {author} {\bibfnamefont {M.~F.}\ \bibnamefont {Hansen}}, \bibinfo {author} {\bibfnamefont {C.~B.}\ \bibnamefont {Koch}}, \bibinfo {author} {\bibfnamefont {K.}~\bibnamefont {Lefmann}},\ and\ \bibinfo {author} {\bibfnamefont {S.}~\bibnamefont {Mørup}},\ }\bibfield  {title} {\bibinfo {title} {Magnetic properties of hematite nanoparticles},\ }\href {https://doi.org/10.1103/PhysRevB.61.6826} {\bibfield  {journal} {\bibinfo  {journal} {Phys. Rev. B}\ }\textbf {\bibinfo {volume} {61}},\ \bibinfo {pages} {6826} (\bibinfo {year} {2000})}\BibitemShut {NoStop}%
\bibitem [{\citenamefont {Fischer}\ \emph {et~al.}(2020)\citenamefont {Fischer}, \citenamefont {Althammer}, \citenamefont {Vlietstra}, \citenamefont {Huebl}, \citenamefont {Goennenwein}, \citenamefont {Gross}, \citenamefont {Gepr\"ags},\ and\ \citenamefont {Opel}}]{Fischer2020}%
  \BibitemOpen
  \bibfield  {author} {\bibinfo {author} {\bibfnamefont {J.}~\bibnamefont {Fischer}}, \bibinfo {author} {\bibfnamefont {M.}~\bibnamefont {Althammer}}, \bibinfo {author} {\bibfnamefont {N.}~\bibnamefont {Vlietstra}}, \bibinfo {author} {\bibfnamefont {H.}~\bibnamefont {Huebl}}, \bibinfo {author} {\bibfnamefont {S.~T.}\ \bibnamefont {Goennenwein}}, \bibinfo {author} {\bibfnamefont {R.}~\bibnamefont {Gross}}, \bibinfo {author} {\bibfnamefont {S.}~\bibnamefont {Gepr\"ags}},\ and\ \bibinfo {author} {\bibfnamefont {M.}~\bibnamefont {Opel}},\ }\bibfield  {title} {\bibinfo {title} {Large spin {H}all magnetoresistance in antiferromagnetic $\alpha$-{Fe}$_2${O}$_3$/{Pt} heterostructures},\ }\href {https://doi.org/10.1103/PhysRevApplied.13.014019} {\bibfield  {journal} {\bibinfo  {journal} {Phys. Rev. Appl.}\ }\textbf {\bibinfo {volume} {13}},\ \bibinfo {pages} {014019} (\bibinfo {year} {2020})}\BibitemShut {NoStop}%
\bibitem [{\citenamefont {Kanj}\ \emph {et~al.}(2023)\citenamefont {Kanj}, \citenamefont {Gomonay}, \citenamefont {Boventer}, \citenamefont {Bortolotti}, \citenamefont {Cros}, \citenamefont {Anane},\ and\ \citenamefont {Lebrun}}]{Kanj2023}%
  \BibitemOpen
  \bibfield  {author} {\bibinfo {author} {\bibfnamefont {A.~E.}\ \bibnamefont {Kanj}}, \bibinfo {author} {\bibfnamefont {O.}~\bibnamefont {Gomonay}}, \bibinfo {author} {\bibfnamefont {I.}~\bibnamefont {Boventer}}, \bibinfo {author} {\bibfnamefont {P.}~\bibnamefont {Bortolotti}}, \bibinfo {author} {\bibfnamefont {V.}~\bibnamefont {Cros}}, \bibinfo {author} {\bibfnamefont {A.}~\bibnamefont {Anane}},\ and\ \bibinfo {author} {\bibfnamefont {R.}~\bibnamefont {Lebrun}},\ }\bibfield  {title} {\bibinfo {title} {Antiferromagnetic magnon spintronic based on nonreciprocal and nondegenerated ultra-fast spin-waves in the canted antiferromagnet $\alpha$-{Fe$_2$O$_3$}},\ }\href {https://doi.org/10.1126/sciadv.adh1601} {\bibfield  {journal} {\bibinfo  {journal} {Sci. Adv.}\ }\textbf {\bibinfo {volume} {9}},\ \bibinfo {pages} {eadh1601} (\bibinfo {year} {2023})}\BibitemShut {NoStop}%
\bibitem [{\citenamefont {Lebrun}\ \emph {et~al.}(2020)\citenamefont {Lebrun}, \citenamefont {Ross}, \citenamefont {Gomonay}, \citenamefont {Baltz}, \citenamefont {Ebels}, \citenamefont {Barra}, \citenamefont {Qaiumzadeh}, \citenamefont {Brataas}, \citenamefont {Sinova},\ and\ \citenamefont {Kl{\"a}ui}}]{Lebrun2020}%
  \BibitemOpen
  \bibfield  {author} {\bibinfo {author} {\bibfnamefont {R.}~\bibnamefont {Lebrun}}, \bibinfo {author} {\bibfnamefont {A.}~\bibnamefont {Ross}}, \bibinfo {author} {\bibfnamefont {O.}~\bibnamefont {Gomonay}}, \bibinfo {author} {\bibfnamefont {V.}~\bibnamefont {Baltz}}, \bibinfo {author} {\bibfnamefont {U.}~\bibnamefont {Ebels}}, \bibinfo {author} {\bibfnamefont {A.~L.}\ \bibnamefont {Barra}}, \bibinfo {author} {\bibfnamefont {A.}~\bibnamefont {Qaiumzadeh}}, \bibinfo {author} {\bibfnamefont {A.}~\bibnamefont {Brataas}}, \bibinfo {author} {\bibfnamefont {J.}~\bibnamefont {Sinova}},\ and\ \bibinfo {author} {\bibfnamefont {M.}~\bibnamefont {Kl{\"a}ui}},\ }\bibfield  {title} {\bibinfo {title} {Long-distance spin-transport across the morin phase transition up to room temperature in ultra-low damping single crystals of the antiferromagnet $\alpha$-{Fe$_2$O$_3$}},\ }\href {https://doi.org/10.1038/s41467-020-20155-7} {\bibfield  {journal} {\bibinfo  {journal} {Nat. Commun.}\ }\textbf {\bibinfo {volume} {11}},\ \bibinfo
  {pages} {6332} (\bibinfo {year} {2020})}\BibitemShut {NoStop}%
\bibitem [{\citenamefont {Lin}\ \emph {et~al.}(2024)\citenamefont {Lin}, \citenamefont {Chen}, \citenamefont {Lu}, \citenamefont {Liang}, \citenamefont {Feng}, \citenamefont {Yamagami}, \citenamefont {Osiecki}, \citenamefont {Leandersson}, \citenamefont {Thiagarajan}, \citenamefont {Liu}, \citenamefont {Felser},\ and\ \citenamefont {Ma}}]{Lin2024}%
  \BibitemOpen
  \bibfield  {author} {\bibinfo {author} {\bibfnamefont {Z.}~\bibnamefont {Lin}}, \bibinfo {author} {\bibfnamefont {D.}~\bibnamefont {Chen}}, \bibinfo {author} {\bibfnamefont {W.}~\bibnamefont {Lu}}, \bibinfo {author} {\bibfnamefont {X.}~\bibnamefont {Liang}}, \bibinfo {author} {\bibfnamefont {S.}~\bibnamefont {Feng}}, \bibinfo {author} {\bibfnamefont {K.}~\bibnamefont {Yamagami}}, \bibinfo {author} {\bibfnamefont {J.}~\bibnamefont {Osiecki}}, \bibinfo {author} {\bibfnamefont {M.}~\bibnamefont {Leandersson}}, \bibinfo {author} {\bibfnamefont {B.}~\bibnamefont {Thiagarajan}}, \bibinfo {author} {\bibfnamefont {J.}~\bibnamefont {Liu}}, \bibinfo {author} {\bibfnamefont {C.}~\bibnamefont {Felser}},\ and\ \bibinfo {author} {\bibfnamefont {J.}~\bibnamefont {Ma}},\ }\href@noop {} {\bibinfo {title} {Observation of giant spin splitting and d-wave spin texture in room temperature altermagnet {RuO$_2$}}} (\bibinfo {year} {2024}),\ \Eprint {https://arxiv.org/abs/2402.04995} {arXiv:2402.04995 [cond-mat.mtrl-sci]}
  \BibitemShut {NoStop}%
\bibitem [{\citenamefont {McClarty}\ and\ \citenamefont {Rau}(2024)}]{mcclarty_landau_2024}%
  \BibitemOpen
  \bibfield  {author} {\bibinfo {author} {\bibfnamefont {P.~A.}\ \bibnamefont {McClarty}}\ and\ \bibinfo {author} {\bibfnamefont {J.~G.}\ \bibnamefont {Rau}},\ }\bibfield  {title} {\bibinfo {title} {Landau {Theory} of {Altermagnetism}},\ }\href {https://doi.org/10.1103/PhysRevLett.132.176702} {\bibfield  {journal} {\bibinfo  {journal} {Phys. Rev. Lett.}\ }\textbf {\bibinfo {volume} {132}},\ \bibinfo {pages} {176702} (\bibinfo {year} {2024})}\BibitemShut {NoStop}%
\bibitem [{\citenamefont {Aoyama}\ and\ \citenamefont {Ohgushi}(2023)}]{aoyama_piezomagnetic_2023}%
  \BibitemOpen
  \bibfield  {author} {\bibinfo {author} {\bibfnamefont {T.}~\bibnamefont {Aoyama}}\ and\ \bibinfo {author} {\bibfnamefont {K.}~\bibnamefont {Ohgushi}},\ }\href {https://doi.org/10.48550/arXiv.2305.14786} {\bibinfo {title} {Piezomagnetic {Properties} in {Altermagnetic} {MnTe}}} (\bibinfo {year} {2023})\BibitemShut {NoStop}%
\bibitem [{\citenamefont {Urru}\ and\ \citenamefont {Spaldin}(2022)}]{Urru2022}%
  \BibitemOpen
  \bibfield  {author} {\bibinfo {author} {\bibfnamefont {A.}~\bibnamefont {Urru}}\ and\ \bibinfo {author} {\bibfnamefont {N.~A.}\ \bibnamefont {Spaldin}},\ }\bibfield  {title} {\bibinfo {title} {Magnetic octupole tensor decomposition and second-order magnetoelectric effect},\ }\href {https://doi.org/https://doi.org/10.1016/j.aop.2022.168964} {\bibfield  {journal} {\bibinfo  {journal} {Ann. Phys.}\ }\textbf {\bibinfo {volume} {447}},\ \bibinfo {pages} {168964} (\bibinfo {year} {2022})}\BibitemShut {NoStop}%
\end{thebibliography}%
\end{document}